\documentclass[prd,amsmath,amssymb,superscriptaddress,floatfix,nofootinbib,10pt]{revtex4}
\usepackage{times}
\usepackage{amssymb,amsbsy,amsmath,amsfonts}
\usepackage{graphicx}
\usepackage{float}
\usepackage{color}
\usepackage{morefloats}
\usepackage{rotating}
\usepackage{srcltx}
\usepackage{slashed}
\usepackage{multirow}
\usepackage{verbatim}
\usepackage{hyperref}
\usepackage{tabularx}
\usepackage{bm}
\allowdisplaybreaks[4]

\usepackage{bbding}
\usepackage{threeparttable}

\DeclareUnicodeCharacter{2212}{\ensuremath{-}}

\newcommand{\PreserveBackslash}[1]{\let\temp=\\#1\let\\=\temp}
\newcolumntype{C}[1]{>{\PreserveBackslash\centering}p{#1}}
\newcolumntype{R}[1]{>{\PreserveBackslash\raggedleft}p{#1}}
\newcolumntype{L}[1]{>{\PreserveBackslash\raggedright}p{#1}}

\begin{document}

\title{Photoproduction of the $ \Lambda (1800) $}

\begin{abstract}
We have carried out a study of the $\gamma ~ p \rightarrow p K^{+}  K^{*-} (K^{*-} \rightarrow K^{-} \pi^{0}) $, $ \gamma ~p \rightarrow p K^{+}  K^{*-} ( K^{*-} \rightarrow \bar{K}^{0}  \pi^{-} )$ and $ \gamma ~p \rightarrow  K^{+}  K^{-}p$ reactions, producing the $K^+ \Lambda(1800)$ final state,  from the perspective that the $\Lambda(1800)$ resonance is dynamically generated from the interaction of $\bar{K}^* N$ with its coupled vector-baryon channels, in complete analogy to the $\Lambda(1405)$ generated from the interaction of $\bar{K} N$ and its coupled pseudoscalar-baryon channels. The two reactions are complementary and their mass distributions are tied to the particular nature of this resonance in that framework. We provide much information on the shapes and strength of the invariant mass distributions of these reactions, and the energy dependence of the cross sections, that when contrasted with future experiments should shed valuable light on the nature of this resonance and its analogy to the $\Lambda(1405)$. 
 
\end{abstract}

%\pacs{13.75.Ev,12.39.Fe,21.30.Fe}
%\keywords{}

\date{\today}
\date{\today}

\author{Melahat Bayar}
\email[]{melahat.bayar@kocaeli.edu.tr}
\affiliation{Department of Physics, Kocaeli Univeristy, 41380, Izmit, Turkey}

\affiliation{Departamento de Física Teórica and IFIC, Centro Mixto Universidad de Valencia-CSIC Institutos de Investigación de Paterna, 46071 Valencia, Spain}

\author{Jing Song}
\email[]{Song-Jing@buaa.edu.cn}
\affiliation{School of Physics, Beihang University, Beijing, 102206, China}
\affiliation{Departamento de Física Teórica and IFIC, Centro Mixto Universidad de Valencia-CSIC Institutos de Investigación de Paterna, 46071 Valencia, Spain}

\author{L. R. Dai}
\email[]{ dailianrong@zjhu.edu.cn}
\affiliation{School of science, Huzhou University, Huzhou, 313000, Zhejiang, China}
\affiliation{Departamento de Física Teórica and IFIC, Centro Mixto Universidad de Valencia-CSIC Institutos de Investigación de Paterna,  46071 Valencia, Spain}

\author{ Eulogio Oset}
\email[]{oset@ific.uv.es}
\affiliation{Departamento de Física Teórica and IFIC, Centro Mixto Universidad de Valencia-CSIC Institutos de Investigación de Paterna, 46071 Valencia, Spain}

\maketitle

\section{introduction}

Hyperons, as baryons containing one or more strange quarks, have attracted much attention in hadron physics, both experimentally~\cite{SAPHIR:1998fev,Tovee:1971ga,Ciborowski:1982et,E761:1993qya,Duryea:1991ck,Anisovich:2007bq,Wilkinson:1981jy,CLAS:2016wrl,delaVaissiere:1984xg,SAPHIR:1999wfu,BESIII:2020uqk,NA61SHINE:2015haq,Biagi:1981cu,Dobbs:2014ifa,ALICE:2017pgw,Chien:1966tr,Sarantsev:2019xxm,Carmony:1964zza,Beretvas:1986km,Bardadin-Otwinowska:1974lrc,Ronniger:2011td,HADES:2017njk,Crede:2023ncq,PANDA:2023ljx,Arellano:2024fym} and theoretically~\cite{Nakamura:2013boa,VanCauteren:2003hn,Jackson:2015dva,Bernard:1992xi,Lee:1968ehl,Feldman:1961su,Sibirtsev:2006uy,Ellis:2007ig,Kim:2012pz,Kaiser:2005tu,Ozpineci:2003gm,Zhong:2013oqa,Quigg:1976xs,Shi:2023xfz,Yan:2023yff}.  The Lambda states have been the object of much debate, since from the very beginning, even before the $\Lambda(1405)$ was observed, this resonance had already been predicted as a $\bar{K} N$ bound state in~\cite{Dalitz:1960du,Dalitz:1959dn}. The advent of the chiral unitary approach ~\cite{Kaiser:1995cy,Kaiser:1995eg,Oset:1997it,Oller:2000ma} made this claim more transparent, and using a unitary approach in the coupled channels to $\bar{K} N$ with the input of the potentials from Chiral Lagrangians~\cite{Ecker:1994gg,Bernard:1995dp}, two states of the $\Lambda(1405)$ were generated~\cite{Oller:2000fj,Jido:2003cb}  that are now reported in the PDG~\cite{ParticleDataGroup:2024cfk}.
These states stem from the interaction of pseudoscalar mesons with the baryons of the octet. The extension of the idea to the interaction of vector mesons with the octet of baryons was done in Ref.~\cite{Oset:2010tof}, where also many dynamically generated states were obtained. In analogy to the $\Lambda(1420)$, and another $\Lambda$, the $\Lambda(1380)$, which couple mostly to $\bar{K} N$ and $\pi \Sigma$, respectively, two analogous states were generated, the $\Lambda(1800)$ coupling mostly to  $\bar{K}^* N$  and the $\Lambda(1900)$ coupling mostly to $\rho \Sigma$. It might look surprising that now the $\bar{K}^* N$ state has smaller mass than the $\rho \Sigma$, compared to the $\bar{K} N$ and $\pi \Sigma$, but the large mass of the $\rho$ compared to that of the pion is responsible for it. 

  The analogy of the $\Lambda(1800)$ to the $\Lambda(1420)$ and the large amount of work devoted to the study of the nature of the two $\Lambda(1405)$ states~\cite{Lutz:2001yb,Garcia-Recio:2002yxy,Magas:2005vu,Ikeda:2012au,Guo:2012vv,Mai:2014xna,Roca:2013av,Roca:2013cca,Cieply:2016jby,Cieply:2011nq,Kamiya:2016jqc,Hyodo:2007jq,Revai:2017isg,Bruns:2019bwg,Miyahara:2018lud,Hyodo:2011ur,Meissner:2020khl}, justifies to turn now the attention to the $\Lambda(1800)$. 
  
  Apart from data of $\bar{K} N$ going to many final states, mostly used to study the $\bar{K} N$ interaction and the related dynamically generated states, the photoproduction reaction, $\gamma p \to K \pi \Sigma$~\cite{CLAS:2013rjt,CLAS:2013rxx,Schumacher:2013vma}, turned out to be a rich source of information on that issue, as shown in Refs.~\cite{Mai:2014xna,Roca:2013av,Roca:2013cca}. From this perspective, our purpose in this work is to study theoretically the $\gamma p\to K \bar{K} N$ and 
$\gamma p\to K \bar{K}^* N$ reactions, with the aim of learning about the nature of this $\Lambda(1800)$ state. The reactions proposed can be easily implemented in present facilities, like Jefferson Lab.

%\newpage
\section{Formalism}
To investigate the photoproduction of the $\Lambda(1800)$, we will employ two mechanisms.

\subsection{Formalism for the photoproduction  of $\gamma ~ p \rightarrow p K^{+}  K^{*-} (K^{*-} \rightarrow K^{-} \pi^{0}) $ and $ \gamma ~p \rightarrow p K^{+}  K^{*-} ( K^{*-} \rightarrow \bar{K}^{0}  \pi^{-} )$  Reactions}

\begin{figure}[h!]
  \centering
  \includegraphics[width=0.50\textwidth]{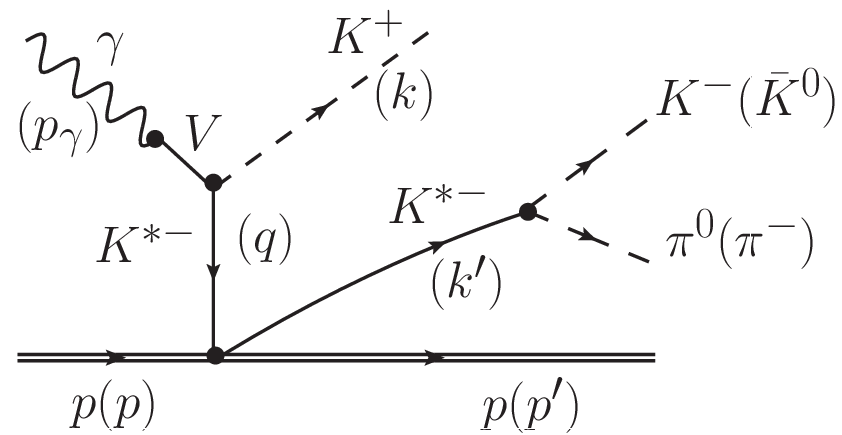}
  \caption {Mechanism of the photoproduction reaction  $\gamma ~ p \rightarrow p K^{+}  K^{*-} (K^{*-} \rightarrow K^{-} \pi^{0}) $ and $ \gamma ~p \rightarrow p K^{+}  K^{*-} ( K^{*-} \rightarrow \bar{K}^{0}  \pi^{-} )$. The symbol $ V $ stands for the $ \rho^{0} $, $ \omega $ and $ \phi $ vector mesons. }
   \label{feynDiag1}
\end{figure}

The mechanism for the  $\gamma ~ p \rightarrow p K^{+}  K^{*-} (K^{*-} \rightarrow K^{-} \pi^{0}) $ and $ \gamma ~p \rightarrow p K^{+}  K^{*-} ( K^{*-} \rightarrow \bar{K}^{0}  \pi^{-} )$  process is shown in Fig. \ref{feynDiag1}. First, we calculate the $\gamma \rightarrow K^{+} K^{*-}$ vertex. As depicted in Fig. \ref{feynDiag1}, we rely upon vector meson dominance, where the photon converts  into a vector meson $V (\rho^{0}, \omega, \phi)$, which subsequently interacts with $K^{+}$ and $K^{*-}$ mesons. The photon conversion into $V$ is derived from the Local Hidden Gauge Lagrangian \cite{BandoLHG1, BandoLHG2, UlfLHG} as described in \cite{NagahiroVDM}:
\begin{align}
{\cal L}_{\gamma V}=-M^{2}_{V}\frac{e}{g}A_{\mu} \langle V^\mu Q\rangle \label{eq:LAGVMD} 
\end{align}
where  $ A_{\mu}  $ and $ V^\mu $  represent the photon and vector meson fields, respectively, and $ Q=1/3(2,-1,-1) $ is the charge
matrix of the $u$, $d$, and $s$ quarks.  In Eq. (\ref{eq:LAGVMD}), $ g $ is the universal coupling in the Local Hidden Gauge Lagrangian, given by $ g=M_{V}/(2 f_{\pi})$, where $ M_{V} $  is an appropriate vector meson
mass (we take $800$ MeV)  and  $ f_{\pi} =93 $ MeV is the pion decay constant. The constant $e$ is the charge of the electron, normalized such that $e^{2}/(4 \pi) = \alpha = 1/137$. 

The other ingredient we need is the $ V(\rho^{0}, \omega, \phi) K^{+} K^{*-} $ vertex, which is described by the anomalous Lagrangian \cite{Bramon92}:
\begin{align}
{\cal L}_{VVP}=\frac{G'}{\sqrt{2}} \epsilon^{\mu \nu \alpha \beta} \langle \partial_{\mu} V_{\nu} \partial_{\alpha} V_{\beta} P \rangle \label{eq:VVP} 
\end{align}
where $  V $ is the matrix of vector meson nonet SU (3) matrices:
\begin{equation}\label{Vecmatrix}
V = \begin{pmatrix}
 \frac{1}{\sqrt{2}}\rho^0 + \frac{1}{\sqrt{2}} \omega & \rho^+ & K^{* +} &  \\
 \rho^- & -\frac{1}{\sqrt{2}}\rho^0 + \frac{1}{\sqrt{2}} \omega  & K^{* 0}  \\
 K^{* -} & \bar{K}^{* 0}  & \phi
\end{pmatrix},
\end{equation}
and $ P $ is the matrix of pseudoscalar fields:
\begin{equation}\label{Psematrix}
P = \begin{pmatrix}
\frac{1}{\sqrt{2}}\pi^0 + \frac{1}{\sqrt{3}} \eta + \frac{1}{\sqrt{6}}\eta' & \pi^+ & K^+ &  \\
 \pi^- & -\frac{1}{\sqrt{2}}\pi^0 + \frac{1}{\sqrt{3}} \eta + \frac{1}{\sqrt{6}}\eta' & K^0  \\
 K^- & \bar{K}^0 & -\frac{1}{\sqrt{3}} \eta + \sqrt{\frac{2}{3}}\eta' 
\end{pmatrix},
\end{equation}
with the $ \eta $-$ \eta' $ mixing of Ref. \cite{Bramon92}, with the symbol $\langle ~ \rangle $ representing the trace in $SU (3)$  space. In Eq. (\ref{eq:VVP}), $ G' $ is the coupling constant of the $ VVP $ Lagrangian which is given by:

\begin{align}
G'=\dfrac{3 g'^{2}}{4 \pi^{2} f} ~~;~~~ g'=-\dfrac{G_{V} M_{V}}{\sqrt{2} f^{2}}. \label{eq:Gpr} 
\end{align}
and $ G_{V}  $ is the coupling of $ \rho $ to two pions, which, as in \cite{Oset:2002sh,Oset:2008hp} we take as  $ G_{V} =69 $ MeV {$(G’=0.0154 ~\mathrm{MeV}^{-1}$) and $G_V= 55$ MeV ($G’=0.0098~ \mathrm{MeV}^{-1}$). We take the average  $G' = 0.012$ $\mathrm{MeV}^{-1}$}.

Using Eqs. (\ref{eq:VVP}, \ref{Vecmatrix}, \ref{Psematrix}), the $V K^{*-} K^{+}$ vertex can be evaluated as follows:
\begin{align}
-i~t = i~\frac{G'}{\sqrt{2}} \frac{1}{3}  \frac{e}{g} \epsilon^{ \mu \nu \alpha \beta}(-i~p_{\gamma})_{ \mu}\epsilon_{ \nu}(\gamma)(i~q_{\alpha})\epsilon_{ \beta}(K^{*-})
 \label{itp8}
\end{align}
We choose to evaluate this amplitude in the $K^{*-} p$ rest frame where the amplitude is in $S$-wave, which leads us to take $\epsilon^{\mu \nu \alpha \beta} q ^\alpha \to \epsilon^{ \mu \nu 0 \beta} q^0$. 
%For the photon, since $\epsilon_{ \nu}$ is spatial, we work in the Coulomb Gauge, $\epsilon^{0}(0)=0$ and the  photon are transverse. 
For the photon, we work in the Coulomb Gauge,  $\epsilon^{0}(0)=0$,  and the photons have only the transverse polarizations. 
With these considerations, we obtain the following result:
\begin{align}
-i~t &= i~\frac{G'}{\sqrt{2}} \frac{1}{3}  \frac{e}{g} \epsilon^{ \mu \nu 0 \beta}p_{\gamma,\mu}\epsilon_{ \nu}(\gamma)q_0 \epsilon_{ \beta}(K^{*-}) \nonumber \\
 &=-  i~\frac{G'}{\sqrt{2}} \frac{1}{3}  \frac{e}{g} \epsilon^{ ijl}p_{\gamma}^i\epsilon^{ j}(\gamma)q^0 \epsilon^{l}(K^{*-}) 
\label{page9}
\end{align}
Now we introduce the $ K^{*-} $ propagator and the $K^{*-} p \rightarrow K^{*-} p$ amplitude:
\begin{align}
 i~\dfrac{1}{(q^{0})^{2}-\vec{q}^{~2}-m^{2}_{K^{*}}} ~(-i) ~ t_{K^{*-} p \rightarrow K^{*-} p} (M_\mathrm{inv}) ~~\epsilon^{m}(K^{*-} (\mathrm{ingoing})) ~ \epsilon^{m}(K^{*-}(\mathrm{outgoing}))
 \label{propagatorandKsp}
\end{align}
where the $K^{*-} p \rightarrow K^{*-} p$ amplitude is calculated in Ref. \cite{RamosKN} using the Local Hidden Gauge formalism with a coupled-channels unitary approach in the isospin basis. Additionally, Ref. \cite{GarzonKN} includes the box diagram involving the exchange of pseudoscalar mesons. However, we will use a Breit Wigner amplitude to account empirically for the $ \Lambda (1800) $ experimental width, which is larger than predicted by the theory:
\begin{align}
t_{K^{*-} p \rightarrow K^{*-} p}=\dfrac{g_{K^{*-} p}^{2}}{M_\mathrm{inv}(K^{*-} p)-M_{\Lambda(1800)}+i\frac{\Gamma_{\Lambda}(1800)}{2}}. \label{eq:AmpKSp} 
\end{align}
with $M_{\Lambda(1800)}=1809$~MeV and $\Gamma_{\Lambda}(1800)=200$~MeV.
Here, $ g_{K^{*-} p} $ is the coupling constant, which was calculated in Ref. \cite{GarzonKN}  in the isospin basis. Hence the coupling constant to $ K^{*-} p  $ is $ | g_{K^{*-} p} |= \dfrac{1}{\sqrt{2} } | g^{I=0}_{\bar{K}^{*}N} | =2.83$. (Our isospin multiplets are ($K^{*+}$, $ K^{*0}$), ( $\bar{K}^{*0} $ ,$-K^{*-}$)).

For the amplitude corresponding to the mechanism in Fig. \ref{feynDiag1}, 
we average $ |t_{\gamma p \rightarrow K^{+} K^{*-} p}|^2  $ over the spin of the incoming particles and sum over the spin of the outgoing particles. However, since there is no spin dependence on the protons in the $ t_{ K^{*-} p \rightarrow  K^{*-} p } $ amplitude, summing and averaging over the proton polarizations gives us a factor of $1$. Then, we get:

\begin{align}
\overline{\sum} \sum |t_{\gamma p \rightarrow K^{+} K^{*-} p}|^2 = \left( \dfrac{G'}{3\sqrt{2} } \frac{e}{g} q^{0} \right)^{2} \left| \dfrac{1}{(q^{0})^{2}-\vec{q}^{~2}-m^{2}_{K^{*}}} \right|^{2} \left| t_{K^{*-} p \rightarrow K^{*-} p} \right|^{2} \vec{p}^{~2}_{\gamma}
\label{eq:sumsum}
\end{align}
where the magnitudes $q^{0}$ and $p_{\gamma}$ are evaluated in the $K^{*-} p $ rest frame.  However, since the $K^{*-}$  propagator is invariant, it is more convenient to evaluate it in the $\gamma p$ rest frame as:
\begin{align}
\dfrac{1}{(q^{0})^{2}-\vec{q}^{~2}-m^{2}_{K^{*}}} = \dfrac{1}{-2 p_{\gamma/\mathrm{c.m}} [\omega (k)-k cos\theta ]+m^{2}_{K}-m^{2}_{K^{*}}- i m_{K^{*}} \Gamma_{K^{*}} }.
\label{eq:KSpropagator}
\end{align}
Here, $  p_{\gamma/\mathrm{c.m}} = (s-m^{2}_{p})/2 \sqrt{s} $  represents the photon momentum in the $ \gamma p $ rest frame, and $K$ and $\theta$ are the kaon momentum and the angle between the kaon and the proton in that frame. We  multiply the $ K^{*-} $ propagator by a form factor, $ F(q) = \Lambda^{2} /(\Lambda^{2} + \vec{q}^{~2} ) $, where $  \Lambda =1000$ MeV. 

 The magnitudes  $q^{0}$ and $p_{\gamma}$ in Eq (\ref{eq:sumsum}), have to be then evaluated in the  $K^{*−}p$ rest frame and they are given by
\begin{align}\label{18}
p_{\gamma}=\dfrac{ p_{\gamma/\mathrm{c.m}}}{M_\mathrm{inv}}
\left\lbrace E_{p}(p_{\gamma/\mathrm{c.m}})+p_{\gamma/\mathrm{c.m}}-\omega (k)+k~cos\theta \right\rbrace 
\end{align}
\begin{align}\label{19}
q^{0}=\dfrac{ 1}{M_\mathrm{inv}}
\left\lbrace [p_{\gamma/\mathrm{c.m}}- \omega (k)] [\sqrt{s}-\omega (k)]+p_{\gamma/\mathrm{c.m}}  k cos\theta -k^{2}\right\rbrace.
\end{align}

Finally, to evaluate the phase space of the cross section, we take the photon in the $ z $ direction.  Using the Mandl and Shaw normalization \cite{Mandl}, the cross section is given by:
\begin{align}
\sigma &= \dfrac{4 m^{2}_{p}}{2 (s-m^{2}_{p})}  \int \dfrac{d^{3} k}{(2 \pi)^{3}} \dfrac{1}{2 \omega (k)} \,
 \int \dfrac{d^{3} k'}{(2 \pi)^{3}} \dfrac{1}{2 \omega (k')} \,
\int \dfrac{d^{3} p'}{(2 \pi)^{3}} \dfrac{1}{2 E(p')} \,
\nonumber \\
&\times
\overline{\sum} \sum |t_{\gamma p \rightarrow K^{+} K^{*-} p}|^2 ~ (2 \pi)^{4}~ \delta^{4}(p_{\gamma}+p-k-k'-p')
\label{eq:sig1}
\end{align}

Taking into account that $ d^{3} q /2 \omega (q) $ is a Lorentz invariant,
we evaluate the $ k' $ and $ p' $ integrals in the frame where $ \vec{p}_{\gamma}+\vec{p}-\vec{k}=0 $.  Using the same argument, the $ k $ integral  is then evaluated in the $ \gamma p $ rest frame. Then the cross section is:
\begin{align}
\sigma = \dfrac{2 m^{2}_{p}}{(s-m^{2}_{p})} \dfrac{1}{32 \pi^{3}} \frac{1}{\sqrt{s}} 
\int_{M_\mathrm{inv/min}}^{M_\mathrm{inv/max}} dM_\mathrm{inv} (K^{*-} p) \,
\int_{-1}^{+1} d(cos\theta)  \, \overline{\sum} \sum |t_{\gamma p \rightarrow K^{+} K^{*-} p}|^2 k ~k'
\label{eq:sig2}
\end{align}
where $ M_\mathrm{inv} (K^{*-} p) $ is the invariant mass of the $ K^{*-} p $ system and $ s $ the ordinary Mandelstam variable for the center-of-mass (c.m.) energy of the $ \gamma p $ initial system. As mentioned before, the angle $ \theta $ is the angle between the $ K^{+} $ and the photon in the $ \gamma p $ rest frame. The momenta $ k $ and $ k' $  are obtained as:
\begin{align}
&\omega (k) =\dfrac{s+m^{2}_{K}-M^{2}_\mathrm{inv} }{2 \sqrt{s}};~~~~~~~~~~k=\sqrt{\omega (k)^{2} -m^{2}_{K}}, \nonumber \\
& k'= \dfrac{\lambda^{1/2} (M^{2}_\mathrm{inv},m^{2}_{K^{*}},m^{2}_{p})}{2 M_\mathrm{inv}},
\label{eq:kandkpr}
\end{align}
where {$M_\mathrm{inv}$ stands for $ M_\mathrm{inv} (K^{*-} p) $} and  $  \lambda$  is the Källén function.

 Finally, since the  $ K^{*-} $ is an unstable particle, with a width, or equivalently, it has a mass distribution, we must then consider this mass distribution given by the spectral function. {The decay channels are $  \pi K $ and we choose the $ \pi^{-} \bar{K}^{0}$ channel,} for its observation. Then we have 

\begin{align}
\dfrac{d\sigma}{d \tilde{M}_{K^{*}}}=-\frac{1}{\pi} 2 \tilde{M}_{K^{*}} \texttt{Im} \dfrac{1}{\tilde{M}^{2}_{K^{*}} -M_{K^{*}}^{2} +i \tilde{M}_{K^{*}} \Gamma (\tilde{M}_{K^{*}})} \sigma(\tilde{M}_{K^{*}})
\label{eq:dGdM}
\end{align}
where $ \tilde{M}_{K^{*}} $ is equal to $ M_\mathrm{inv} (\pi^{-} \bar{K}^{0})$ and $ \Gamma (\tilde{M}_{K^{*}} ) $    is given by;
\begin{align}
\Gamma (\tilde{M}_{K^{*}} )= \dfrac{M_{K^{*}}^{2}}{\tilde{M}^{2}_{K^{*}}} \left( \dfrac{p_{\pi}}{p_{\pi,\mathrm{on}}} \right)^{3} \Gamma_\mathrm{on}
\end{align}
with $ \Gamma_\mathrm{on} =51.4$~MeV, the $ K^{*-}  $ width, and 
\begin{align}
 p_{\pi}= \dfrac{\lambda^{1/2} (\tilde{M}^{2}_{K^{*}},m^{2}_{\pi},m^{2}_{K})}{2 \tilde{M}_{K^{*}}}, ~~~~~~~~~ p_{\pi,\mathrm{on}}= \dfrac{\lambda^{1/2} (M^{2}_{K^{*}},m^{2}_{\pi},m^{2}_{K})}{2 M_{K^{*}}}.
\end{align}
Since the $ K^{*} $ has an isospin of $ \frac{1}{2} $, searching for the $ K^{*-} $ in the $ \pi^{-} \bar{K}^{0} $ channel,   we have
\begin{align}
\vert  \pi K; \frac{1}{2},-\frac{1}{2} \rangle =-\frac{1}{\sqrt{3}} \vert  \pi^{0} K^{-}  \rangle -\sqrt{\frac{2}{3}} \vert  \pi^{-} \bar{K}^{0}  \rangle
\label{eq:iso}
\end{align}
then we must multiply the Eq. (\ref{eq:dGdM}) by  $ 2/3 $ if we select the $\pi^{-} \bar{K}^{0} $ decay mode. 

Thus, finally, we get, for  $\pi^{-} \bar{K}^{0} $ final decay product of the $ K^{*-} $,

\begin{align}
\dfrac{d\sigma}{d M_\mathrm{inv} ( \pi^{-} \bar{K}^{0})~~ d M_\mathrm{inv} (  K^{*-} p )}&=-\frac{4}{3 \pi}  M_\mathrm{inv} ( \pi^{-} \bar{K}^{0}) \texttt{Im} \dfrac{1}{  M^{~2}_\mathrm{inv} ( \pi^{-} \bar{K}^{0})-M_{K^{*}}^{2} +i  M_\mathrm{inv} ( \pi^{-} \bar{K}^{0}) \Gamma ( M_\mathrm{inv} ( \pi^{-} \bar{K}^{0}))}  \nonumber \\
& \times\dfrac{2 m^{2}_{p}}{(s-m^{2}_{p})} \dfrac{1}{32 \pi^{3}} \frac{1}{\sqrt{s}} 
\int_{-1}^{+1} d(cos\theta)  \, \overline{\sum} \sum |t_{\gamma p \rightarrow K^{+} K^{*-} p}|^2 k ~k'
\label{eq:dGdMdM}
\end{align}
where $ M_\mathrm{inv} (  K^{*-} p) $ is equal to $ M_\mathrm{inv} (  \pi^{-} \bar{K}^{0} p) $ and $ k'  $ of Eq. (\ref{eq:kandkpr}) is now given by 
\begin{align}
k' = \dfrac{\lambda^{1/2} (M^{2}_\mathrm{inv}(K^{*-} p),M^{2}_\mathrm{inv}(\pi^{-} \bar{K}^{0}),m^{2}_{p})}{2 M_\mathrm{inv}(K^{*-} p)}.
\end{align}
We have conducted the calculations for the $\pi^-\bar{K}^0$ decay mode of the ${K}^{*-}$. Should we have taken instead the $\pi^0K^-$ decay mode, the same results can be used substituting $2/3$ by $1/3$  according to Eq.~(\ref{eq:iso}).

\subsection{Formalism for $\gamma ~ p \rightarrow K^{+}  K^{-}  p $   Reaction and Contact Term}

\begin{figure}[h!]
  \centering
  \includegraphics[width=0.50\textwidth]{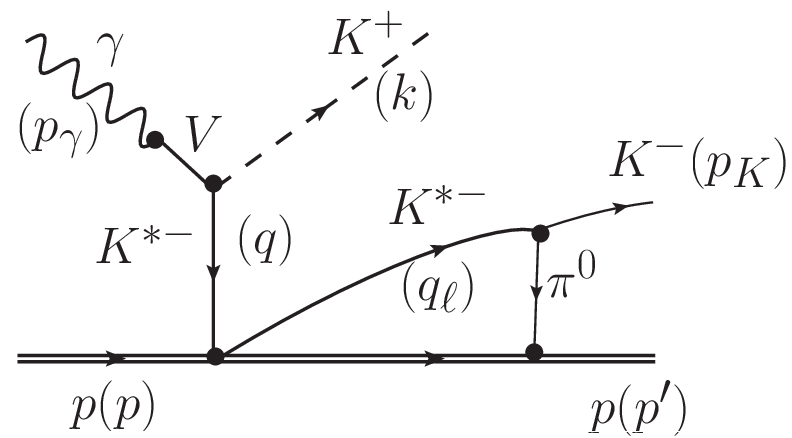}
  \caption {Feynman Diagram of $\gamma ~ p \rightarrow K^{+}  K^{-}  p $. The symbol $ V $ stands for the $ \rho^{0} $, $ \omega $ and $ \phi $ vector mesons. }
   \label{feynDiag2}
\end{figure}

In Fig. \ref{feynDiag2}, we present the Feynman diagram for the $\gamma ~ p \rightarrow K^{+}  K^{-}  p $   reaction.  To calculate this process, we need to evaluate two additional vertices that differ from those in Fig. \ref{feynDiag1}: the upper vertex, $ K^{*-} \rightarrow K^{-}  \pi^{0}$, and the lower vertex, $ \pi^{0} p p $

The Lagrangian for the upper vertex  ($VPP$ vertex) is given by
\begin{align}
 {\cal L}_{VPP}&=-ig\langle [P,\partial_\mu P]V^\mu \rangle \label{eq:LAG} \\ g&=\dfrac{M_V}{2 f} ~(M_V\simeq800 ~\mathrm{MeV}, f=93 ~\mathrm{MeV})  \nonumber . 
\end{align}
Using the above Lagrangian, one obtains the following expression for the upper vertex:
\begin{align}
-i t =i{\cal L}=-i g \frac{1}{\sqrt{2}} \vec{\epsilon}~(K^{*}) (\vec{p}_{K}- \vec{p}_{\pi}) \nonumber . 
\end{align}
where $ \vec{p}_{K} $ represents the momentum of the outgoing  $ K^{-} $ meson, and $ \vec{p}_{\pi} $ the momentum of the intermediate pion.  For the lower vertex, we apply the results from Ref. \citep{GarzonKN}:

\begin{align}
-i {t}= \dfrac{D+F}{2 f}    \vec{\sigma}\cdot \vec{p}_\pi 
\end{align}
with $ D=0.75 $ and $ F=0.51 $. 

The amplitude corresponding to the Feynman Diagram in Fig. \ref{feynDiag2} is given by
\begin{align}
-i~t_{L} &= i~ e \frac{G'}{\sqrt{2}} \frac{1}{3 \sqrt{2}}  \dfrac{D+F}{2 f} \epsilon^{ijm} p^{i}_{\gamma} \epsilon^{j}_{\gamma} q^{0} \dfrac{1}{(q^{0})^{2}-\vec{q}^{~2}-m^{2}_{K^{*}}+i~\epsilon} t_{K^{*-} p,K^{*-} p}(M_\mathrm{inv})  \nonumber \\
 & \times \int\frac{d^4q_{\ell}}{(2\pi)^4}  \dfrac{1}{q_{\ell}^{2}-m^{2}_{K^{*}}+i~\epsilon} \dfrac{M_{p}}{E(\vec{P}-\vec{q_{\ell}})} \dfrac{1}{P^{0}- q^{0}_{\ell}-E(\vec{P}-\vec{q_{\ell}})+i~\epsilon}  \nonumber \\
 & \times \dfrac{1}{p_{\pi}^{2}-m^{2}_{\pi}+i~\epsilon}  \vec{\sigma} \cdot \vec{p}_{\pi} ~ ( \vec{p}_{K} -\vec{p}_{\pi})^{m}
\label{eq:t4ddiag2}
\end{align}
{After performing the $ q_l^{0} $ integration analytically, we obtain }
\begin{align}
-i~t_{L} &=  e \frac{G'}{\sqrt{2}} \frac{1}{3 \sqrt{2}}  \dfrac{D+F}{2 f} q^{0} \dfrac{1}{(q^{0})^{2}-\vec{q}^{~2}-m^{2}_{K^{*}}+i~\epsilon} t_{K^{*-} p,K^{*-} p}(M_\mathrm{inv})  \nonumber \\
 &\times \epsilon^{ijm} p^{i}_{\gamma} \epsilon^{j}_{\gamma} \sigma^{r} \left[ A ~p^{2}_{K}~ \delta_{rm}+ B~ p_{K,r} ~p_{K,m} \right] 
\label{eq:t3ddiag2}
\end{align}
where
\begin{align}
A   &= \frac{1}{{2}}  \int\frac{d^3\vec{q}_{\ell}}{(2\pi)^3} \frac{1}{2\omega_{\pi}( \vec{q}_{\ell}-\vec{p}_{K})}  \frac{1}{2\omega_{K^*}( \vec{q}_{\ell})} 
\dfrac{M_{p}}{E(\vec{q_{\ell}})}
 \dfrac{1}{M_\mathrm{inv} - E(\vec{q}_{\ell}) - \omega_{K^*}( \vec{q}_{\ell}) + i \dfrac{\Gamma_{K^{*}} }{2} }  \theta(q_{\rm max}-| \vec{q}_{\ell}|)   \nonumber \\
& \times  \left\lbrace \dfrac{1}{M_\mathrm{inv} - E(\vec{q}_{\ell}) - p_{K}^{0} - \omega_{\pi}( \vec{q}_{\ell} -\vec{p}_{K}) + i \epsilon } 
+ \dfrac{1}{ p_{K}^{0} -  \omega_{K^*}( \vec{q}_{\ell}) - \omega_{\pi}( \vec{q}_{\ell} -\vec{p}_{K})  + i \dfrac{\Gamma_{K^{*}} }{2}} 
\right\rbrace 
\nonumber \\
& \times  \left\lbrace \dfrac{1}{\vec{p}^{~2}_{K}} ~ (\vec{q}_{\ell} - \vec{p}_{K})  \cdot ( 2 \vec{p}_{K} - \vec{q}_{\ell} )
- \dfrac{1}{\vec{p}^{~4}_{K}} \left[ (\vec{q}_{\ell} - \vec{p}_{K})  \cdot \vec{p}_{K} \right]   \left[ ( 2 \vec{p}_{K} - \vec{q}_{\ell} )  \cdot \vec{p}_{K} \right] 
\right\rbrace 
\label{eq:AA}
\end{align}
and 
\begin{align}
B   &=- \frac{1}{{2}}  \int\frac{d^3\vec{q}_{\ell}}{(2\pi)^3} \frac{1}{2\omega_{\pi}( \vec{q}_{\ell}-\vec{p}_{K})}  \frac{1}{2\omega_{K^*}( \vec{q}_{\ell})} 
\dfrac{M_{p}}{E(\vec{q_{\ell}})}
 \dfrac{1}{M_\mathrm{inv} - E(\vec{q}_{\ell}) - \omega_{K^*}( \vec{q}_{\ell}) + i \dfrac{\Gamma_{K^{*}} }{2} }  \theta(q_{\rm max}-| \vec{q}_{\ell}|)   \nonumber \\
& \times  \left\lbrace \dfrac{1}{M_\mathrm{inv} - E(\vec{q}_{\ell}) - p_{K}^{0} - \omega_{\pi}( \vec{q}_{\ell} -\vec{p}_{K}) + i \epsilon } 
+ \dfrac{1}{ p_{K}^{0} -  \omega_{K^*}( \vec{q}_{\ell}) - \omega_{\pi}( \vec{q}_{\ell} -\vec{p}_{K})  + i \dfrac{\Gamma_{K^{*}} }{2}} 
\right\rbrace 
\nonumber \\
& \times  \left\lbrace \dfrac{1}{\vec{p}^{~2}_{K}} ~ (\vec{q}_{\ell} - \vec{p}_{K})  \cdot ( 2 \vec{p}_{K} - \vec{q}_{\ell} )
- \dfrac{3}{\vec{p}^{~4}_{K}} \left[ (\vec{q}_{\ell} - \vec{p}_{K})  \cdot \vec{p}_{K} \right]   \left[ ( 2 \vec{p}_{K} - \vec{q}_{\ell} )  \cdot \vec{p}_{K} \right] 
\right\rbrace.
\label{eq:BB}
\end{align}
The factor  $  \theta(q_{\rm max}-| \vec{q}_{\ell}|)   $  in Eqs. (\ref{eq:AA}) and (\ref{eq:BB}) is introduced because the $ t $ matrix $ t_{K^{*-} p,K^{*-} p} (\vec{q}_{\ell}, M_\mathrm{inv} )$ inside a loop factorizes as $  t_{K^{*-} p,K^{*-} p} ( M_\mathrm{inv} )  \theta(q_{\rm max}-| \vec{q}_{\ell}|) $, where $ q_\mathrm{max} $ is the cut off used to regulate the $ G $ function in the study of $V B \rightarrow VB $ scattering matrix ( see Ref. \cite{danijuan}). Here, $ p_{K} $ and $  p_{K}^{0} $ are defined as follows:
\begin{align}
 p_{K} = \dfrac{\lambda^{1/2} (M^{2}_\mathrm{inv}(K^{-} p),m^{2}_{K^{-}},m^{2}_{p})}{2 M_\mathrm{inv}(K^{-} p)}, ~~~~ p_{K}^{0}=\dfrac{M^{2}_\mathrm{inv}(K^{-} p)+m^{2}_{K^{-}}-m^{2}_{p}}{2  M_\mathrm{inv}(K^{-} p)}
\end{align}
\begin{figure}[h!]
  \centering
  \includegraphics[width=0.60\textwidth]{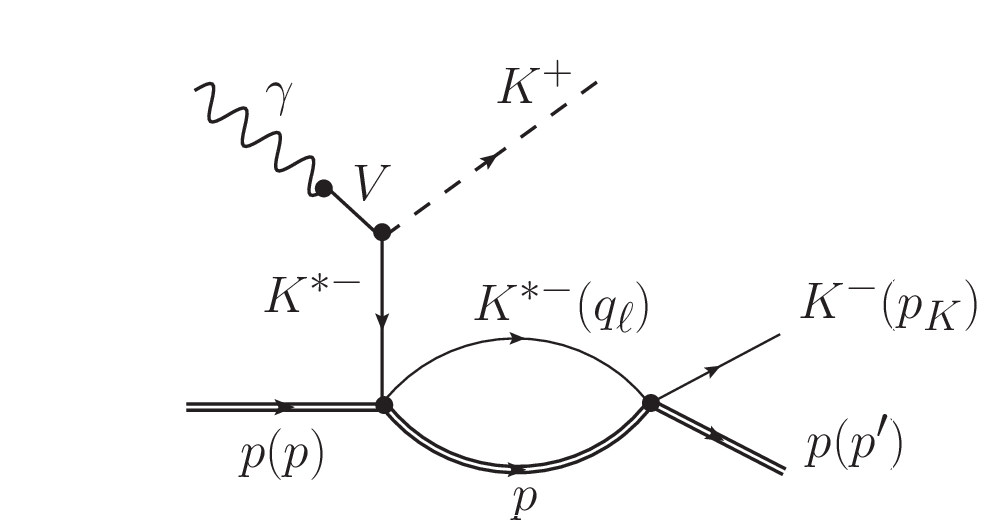}
  \caption {Feynman Diagram of the contact term. }
   \label{feynDiag3}
\end{figure}
To ensure gauge invariance in gamma $ N \rightarrow K N $ when vector meson dominance is used, converting the photon into a vector meson, a contact term has to be introduced~\cite{GarzonKN} in $ K^{*} N \rightarrow K N $. 
This contact term should be added to the amplitudes in Eq. (\ref{eq:t4ddiag2}). The contact term for the   $K^*N \to KN$ amplitude is given by

\begin{align}
-i \tilde{t}_\mathrm{cont}=-g \frac{1}{\sqrt{2}} \dfrac{D+F}{2 f}   \vec{\epsilon}  \cdot \vec{\sigma} 
\end{align}

The corresponding Feynman diagram is shown in Fig. \ref{feynDiag3}, and the matrix element is given by
\begin{align}
-i~t^{(C)}_{L} &= -i~ e \frac{G'}{\sqrt{2}} \frac{1}{3 \sqrt{2}}  \dfrac{D+F}{2 f} \epsilon^{ijm} p^{i}_{\gamma} \epsilon^{j}_{\gamma} q^{0} \sigma^{m} \dfrac{1}{(q^{0})^{2}-\vec{q}^{~2}-m^{2}_{K^{*}}+i~\epsilon} t_{K^{*-} p,K^{*-} p}(M_\mathrm{inv})  \nonumber \\
 & \times  \int\frac{d^4q_{\ell}}{(2\pi)^4}  \dfrac{1}{q_{\ell}^{2}-m^{2}_{K^{*}}+i~\epsilon} \dfrac{M_{p}}{E(\vec{P}-\vec{q_{\ell}})} \dfrac{1}{P^{0}- q^{0}_{\ell}-E(\vec{P}-\vec{q_{\ell}})+i~\epsilon} \theta(q_{\rm max}-| \vec{q}_{\ell}|)  
\label{eq:t4ddiagContact}
\end{align}
As we observe from Eq. (\ref{eq:t4ddiagContact}),  the integral part multiplied by $i$ is a standard $ G $- function of the $ K^{*} $ meson and proton loop. {After performing the $ q_l^{0} $ integration,  it can be expressed in the following form:}
\begin{align}
G_{K^{*} p } (P^{0}) = 2 M_{p} \int_{0}^{q_\mathrm{max}} \frac{d^3\vec{q}_{\ell}}{(2\pi)^3}  \frac{\omega_{K^{*}} + \omega_{p}}{2 \omega_{K^{*}}  \omega_{p}}
\dfrac{1}{P^{0}+\omega_{K^{*}} + \omega_{p} }
\dfrac{1}{P^{0}-\omega_{K^{*}} - \omega_{p}+  i \dfrac{\Gamma_{K^{*}} }{2} } \theta(q_{\rm max}-| \vec{q}_{\ell}|)  
\label{eq:tloopMB}
\end{align}
Here, $ \omega_{K^{*}} =\sqrt{\vec{q}^{~2}_{\ell} + m^{2}_{K^{*}}} $ and  $ \omega_{p} =\sqrt{\vec{q}^{~2}_{\ell} + m^{2}_{p} }$, where  $ q_\mathrm{max} $ is the cut-off for the three-momentum, and $ P^{0} $ represents the center-of-mass (CM) energy.

Hence, our total amplitude is given by
\begin{align}
-i~t^\mathrm{total}_{L} &= -i~(t_{L} + t^{(C)}_{L} )=H  \epsilon^{ijm} p^{i}_{\gamma} \epsilon^{j}_{\gamma} \sigma^{r} \left[ A' ~p^{2}_{K}~ \delta_{rm}+ B~ p_{K,r} ~p_{K,m} \right] 
\label{eq:ttolal}
\end{align}
with $  A' ~p^{2}_{K} \equiv  A ~p^{2}_{K} - G_{K^{*} p } (M_\mathrm{inv}) $  and the coefficient $ H $ is given by:
\begin{align}
H &=  e \frac{G'}{\sqrt{2}} \frac{1}{3 \sqrt{2}}  \dfrac{D+F}{2 f} q^{0} \dfrac{1}{(q^{0})^{2}-\vec{q}^{~2}-m^{2}_{K^{*}}+i~\epsilon} t_{K^{*-} p,K^{*-} p}(M_\mathrm{inv}).  
\end{align}
Finally, for Figs. \ref{feynDiag2} and \ref{feynDiag3}, we obtain: 
\begin{align}
\dfrac{d\sigma}{d M_\mathrm{inv} ( K^{-} p)} = \dfrac{2 m^{2}_{p}}{(s-m^{2}_{p})} \dfrac{1}{32 \pi^{3}} \frac{1}{\sqrt{s}} 
\int_{-1}^{+1} d(cos\theta)  \, \overline{\sum} \sum |t^\mathrm{total}_{L}|^2 k ~k'
\label{eq:dGdMloop}
\end{align}
with 
\begin{align}
\overline{\sum} \sum |t^\mathrm{total}_{L}|^2 =  |H|^2  \vec{p}_{\gamma}^{~2} \vec{p}_{K}^{~4}  \left\lbrace |A'|^2 +\frac{1}{3} [2~ \texttt{Re} (A' B^{*})+ |B|^2] \right\rbrace.
\end{align}
In Eq. (\ref{eq:dGdMloop}), $ k $  and $ k'$ are given by:
\begin{align}
k = \dfrac{\lambda^{1/2} (s,m^{2}_{K^{+}},M^{2}_\mathrm{inv}(K^{-} p)}{2 \sqrt{s}}, ~~~ k' = \dfrac{\lambda^{1/2} (M^{2}_\mathrm{inv}(K^{-} p),m^{2}_{K^{-}},m^{2}_{p})}{2 M_\mathrm{inv}(K^{-} p)}.
\end{align}

\section{Results}
 
In order to find useful information about the $ \Lambda (1800) $ state we have conducted calculations for two reactions. In the first case we produce directly the $ K^{*-} p $ state, which in principle should not show the peak for the  $ \Lambda (1800) $  because it corresponds to a bound state of $ K^{*-} p $. However, due to the width of the $ K^{*-} $ we still can see strength of the reaction below the nominal threshold of $ K^{*-} p $, but reduced by the weight of the $ K^{*-} $ spectral function. The combination of this weight, the phase space and the resonance structure of the $ K^{*-} p \rightarrow  K^{*-} p $ amplitude, have as a consequence that a peak is still seen but displaced at higher energies as we show below.

\begin{figure}[H]
    \centering
\includegraphics[width=0.6\textwidth]{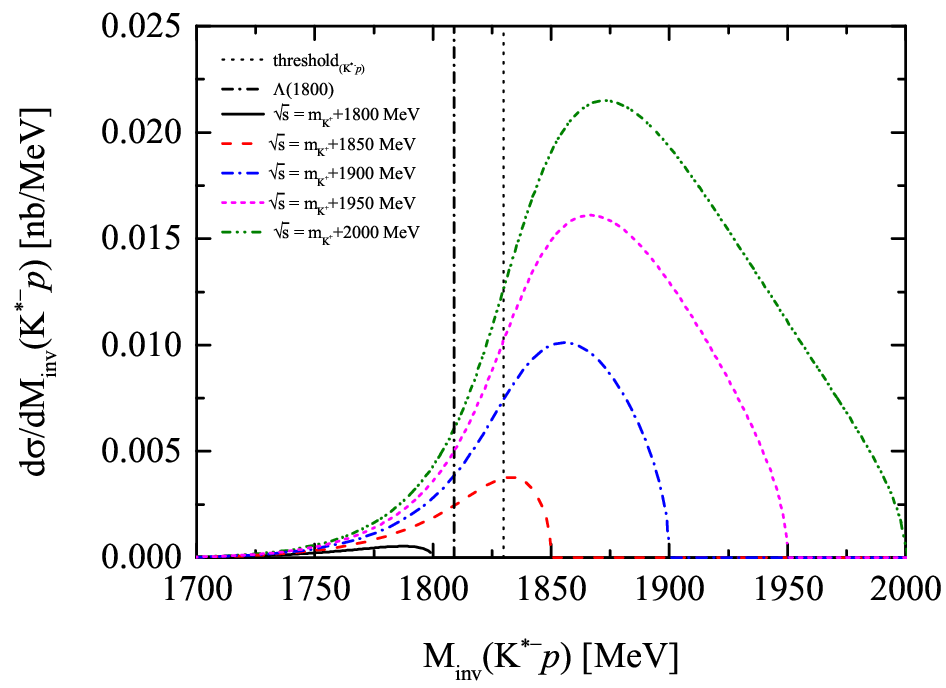}
   \caption{With the integration of the $M_{\mathrm {inv }}(\bar{K}^{0}\pi)$, we present the mass distribution of $(\frac{d\sigma}{dM_{\mathrm {inv }}(K^{*-}p)})$ with the dependence on $M_{\mathrm {inv }}(K^{*-}p)$, The $(\frac{d\sigma}{dM_{\mathrm {inv }}(K^{-}p)})$ values obtained with different    $\sqrt{s}~(\sqrt{s}=m_{K^+}+1800,1850,1900,1950,2000~\mathrm {MeV })$  values are shown by different lines.  And the vertical black dashed line and black dotted line represent the threshold of $M_{K^{*-}p}$ and the boundary of $\Lambda(1800)$, respectively.}
    \label{res1}
\end{figure}

In Fiq. \ref{res1}, we present the results for the mechanism of the $\gamma p \rightarrow K^{+} K^{*-} (K^{*-} \rightarrow \bar{K}^{0} \pi^{-} ) $ process.  To achieve this, we first integrate over $ M_\mathrm{inv} ( \pi^{-} \bar{K}^{0})$ in Eq. (\ref{eq:dGdMdM}), and then we plot $ \dfrac{d\sigma}{d M_\mathrm{inv} (  K^{*-} p )} $  as a function of  $ M_\mathrm{inv} ( K^{*-} p ) $ for different values of $ \sqrt{s} $, ranging from $ \sqrt{s} =m_{K^{-}} + 1800$ MeV to $ \sqrt{s} =m_{K^{-}} + 2000 $ MeV.

%
%\begin{figure}[h!]
%  \centering
  %\includegraphics[width=0.80\textwidth]{tmat1.eps}
%  \caption {Differantial cross section for $\gamma p \rightarrow K^{+} K^{*-} (K^{*-} \rightarrow \bar{K}^{0} \pi^{-} ) $  as a function of the invariant mass of $ K^{*-} p $ for different values of $ \sqrt{s} $. The red-dashed line is the $ K^{*-} p $ threshold.  }
%   \label{res1}
%\end{figure}
%

We can see that we get a peak displaced to higher energies with respect to the nominal mass of the $ \Lambda (1800) $, $ 1800 $ MeV. The peak appears around $ 1870 $ MeV and the width follows approximately the nominal width of $ 200 $ MeV. We also observe that the kinematical factors of the amplitude favor the use of higher $\gamma p $ energies over the small ones. One should also note that by choosing small $ \sqrt{s}  $ one is restricting much the phase space, preventing to see the actual shape of the resonance. 

On the other hand, the choice of the second reaction,  $\gamma p \rightarrow K^{+} K^{-} p $, is made as a complement, not only to see better the peak of the resonance, but also because the strength of the reaction is tied to the consideration of the $ \Lambda (1800) $ as basically a  $ \bar{K}^{*} N $ state. This is something that could be tested in future experiments.

\begin{figure}[H]
    \centering
\includegraphics[width=0.6\textwidth]{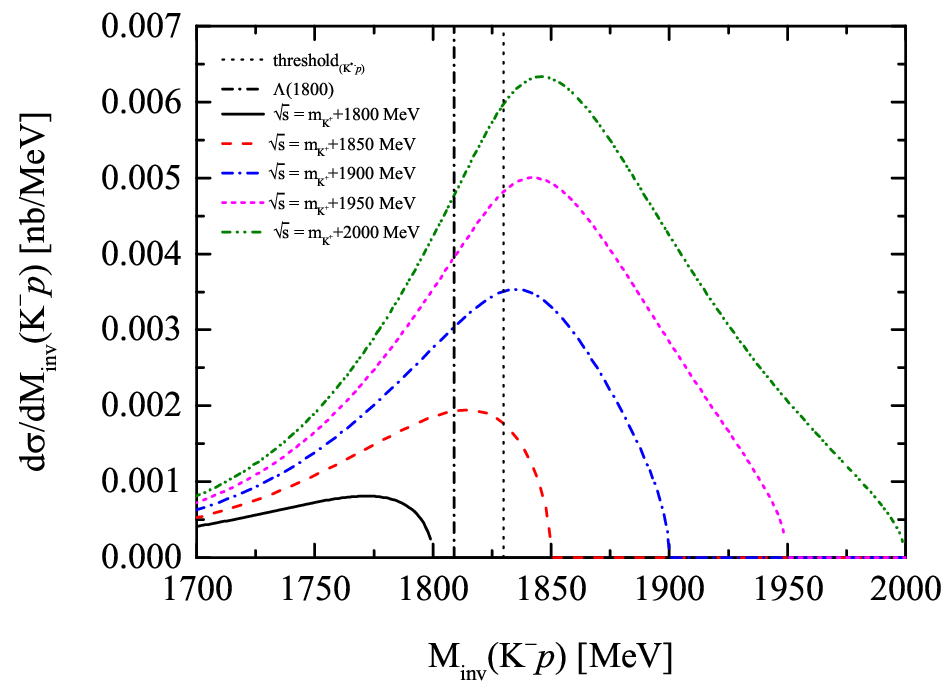}
   \caption{The mass distribution of $(\frac{d\sigma}{dM_{\mathrm {inv }}(K^{-}p)})$ with the dependence on $M_{\mathrm {inv }}(K^{-}p)$.  The labels are same as Fig.~\ref{res1}.}
    \label{res2}
\end{figure}

In Fiq. \ref{res2}, we show the results of Eq. (\ref{eq:dGdMloop}) for the  $\gamma p \rightarrow K^{+} K^{-} p $ reaction for different values of $ \sqrt{s} $, similar to those in Fig. \ref{res1}. 
We see some distinctive feature of this reaction compared to the former one, and it is that now there are no phase space restrictions below the $ K^{*-} p $ threshold and the shape of the resonance is better seen at the higher values of $ \sqrt{s} $, where there are no restrictions on the phase space at high values of $M_\mathrm{inv} (K^{-} p)$. We also observe that the peak now appears at lower energies than in the former reaction, at around $ 1850 $ MeV. The structure of the amplitude and the large width of the resonance have as a consequence this shift of the mass with respect to the nominal one. 

As in the case of the former reaction, we also observe that the strength of the cross section increases with the value of $ \sqrt{s} $. This dependence of the cross section on $ \sqrt{s} $ can also be a testing ground of the dynamics of the reaction based on the molecular assumption for the nature of the $ \Lambda (1800) $ resonance.

%
%\begin{figure}[h!]
%  \centering
  %\includegraphics[width=0.80\textwidth]{tmat2.eps}
%  \caption { $ \dfrac{d\sigma}{d M_\mathrm{inv} ( K^{-} p)} $ as a function of the invariant mass of $K^{-} p $ for different values of $ \sqrt{s} $. The red-dashed line is the $ K^{-} p $ threshold. }
%   \label{res2}
%\end{figure}
%

In summary, our calculations present much information with these two reactions that could be contrasted with experiment when the reactions are actually implemented. The work done here should provide a motivation for the actual measurements in present Laboratories.

%\subsection{2nd reaction: Exchange and contact term}

\section{Conclusions}    
    We have addressed the study of the  $\gamma ~ p \rightarrow p K^{+}  K^{*-} (K^{*-} \rightarrow K^{-} \pi^{0}) $ and $ \gamma ~p \rightarrow p K^{+}  K^{*-} ( K^{*-} \rightarrow \bar{K}^{0}  \pi^{-} )$  plus the $ \gamma ~p \rightarrow p K^{+}  K^{-}p$ 
 reaction in search of information for the $\Lambda(1800)$ resonance from the perspective that this resonance is dynamically generated by the interaction of vector-baryon ($1/2^+$) in the channel of $\bar{K}^*N$ and its coupled channels. In the framework of the chiral unitary approach, where this resonance, among others, is generated, the $\Lambda(1800)$ couples mostly to $\bar{K}^*N$.  From this perspective we have evaluated the cross sections for photoproduction of this resonance, by looking at $K^{+}  K^{*-}p$  production on one side and $K^{+}  K^{-}p$ production on the other. Both reactions are complementary and the strength of the cross sections, the shapes in the invariant mass distributions and the dependence on the total energy of the initial $\gamma p$ system, are tied to the picture of this resonance as a molecular state of $\bar{K}^*N$  with its coupled channels. 

  The two reactions chosen, with $K^{+}  K^{*-}p$ and $K^{+} K^{-}p$ final states are complementary. In the first one, one detects the $K^{*-}$ from its $\pi^- \bar{K}^0$ decay channel, and one can go below the nominal $K^{*-}p$ threshold due to the width of the $K^{*-}$. The combination of the phase space, the resonant pole below the $K^{*-}p$ threshold and the mass distribution of the $K^{*-}$, have as a consequence that a peak is still seen above the $K^{*-}p$ threshold, displaced to higher energies than in the case of the $K^-p$ final state, where there are no restrictions of phase space and the resonance shows up more clearly in the $K^{-}p$  mass distribution, although the peak is also a bit displaced to higher energies with respect the nominal mass of the resonance. 
  
    The features of the two mass distributions, the strength and the $\sqrt{s}$ dependence of the cross sections offer a rich variety of information that can be contrasted with future experiments and should shed light on the nature of this resonance and its analogy to the $\Lambda(1405)$. The results obtained here should provide a motivation to carry on these experiments.   
\section{Acknowledgement} 
This work is partly supported by the National Natural Science
Foundation of China under Grants  No. 12405089, No. 12247108, and No.  12175066 and
the China Postdoctoral Science Foundation under Grant
No. 2022M720360 and No. 2022M720359. This work is also supported by
the Spanish Ministerio de Economia y Competitivi-
dad (MINECO) and European FEDER funds under
Contracts No. FIS2017-84038-C2-1-P B, PID2020-
112777GB-I00, and by Generalitat Valenciana under con-
tract PROMETEO/2020/023. This project has received
funding from the European Union Horizon 2020 research
and innovation programme under the program H2020-
INFRAIA-2018-1, grant agreement No. 824093 of the
STRONG-2020 project.

%\newpage
\bibliography{refs.bib}

\begin{thebibliography}{83}
\expandafter\ifx\csname natexlab\endcsname\relax\def\natexlab#1{#1}\fi
\expandafter\ifx\csname bibnamefont\endcsname\relax
  \def\bibnamefont#1{#1}\fi
\expandafter\ifx\csname bibfnamefont\endcsname\relax
  \def\bibfnamefont#1{#1}\fi
\expandafter\ifx\csname citenamefont\endcsname\relax
  \def\citenamefont#1{#1}\fi
\expandafter\ifx\csname url\endcsname\relax
  \def\url#1{\texttt{#1}}\fi
\expandafter\ifx\csname urlprefix\endcsname\relax\def\urlprefix{URL }\fi
\providecommand{\bibinfo}[2]{#2}
\providecommand{\eprint}[2][]{\url{#2}}

\bibitem[{\citenamefont{Tran et~al.}(1998)}]{SAPHIR:1998fev}
\bibinfo{author}{\bibfnamefont{M.~Q.} \bibnamefont{Tran}} \bibnamefont{et~al.} (\bibinfo{collaboration}{SAPHIR}), \bibinfo{journal}{Phys. Lett. B} \textbf{\bibinfo{volume}{445}}, \bibinfo{pages}{20} (\bibinfo{year}{1998}).

\bibitem[{\citenamefont{Tovee et~al.}(1971)}]{Tovee:1971ga}
\bibinfo{author}{\bibfnamefont{D.~N.} \bibnamefont{Tovee}} \bibnamefont{et~al.}, \bibinfo{journal}{Nucl. Phys. B} \textbf{\bibinfo{volume}{33}}, \bibinfo{pages}{493} (\bibinfo{year}{1971}).

\bibitem[{\citenamefont{Ciborowski et~al.}(1982)}]{Ciborowski:1982et}
\bibinfo{author}{\bibfnamefont{J.}~\bibnamefont{Ciborowski}} \bibnamefont{et~al.}, \bibinfo{journal}{J. Phys. G} \textbf{\bibinfo{volume}{8}}, \bibinfo{pages}{13} (\bibinfo{year}{1982}).

\bibitem[{\citenamefont{Morelos~Pineda et~al.}(1993)}]{E761:1993qya}
\bibinfo{author}{\bibfnamefont{A.}~\bibnamefont{Morelos~Pineda}} \bibnamefont{et~al.} (\bibinfo{collaboration}{E761}), \bibinfo{journal}{Phys. Rev. Lett.} \textbf{\bibinfo{volume}{71}}, \bibinfo{pages}{2172} (\bibinfo{year}{1993}).

\bibitem[{\citenamefont{Duryea et~al.}(1991)}]{Duryea:1991ck}
\bibinfo{author}{\bibfnamefont{J.}~\bibnamefont{Duryea}} \bibnamefont{et~al.}, \bibinfo{journal}{Phys. Rev. Lett.} \textbf{\bibinfo{volume}{67}}, \bibinfo{pages}{1193} (\bibinfo{year}{1991}).

\bibitem[{\citenamefont{Anisovich et~al.}(2007)\citenamefont{Anisovich, Kleber, Klempt, Nikonov, Sarantsev, and Thoma}}]{Anisovich:2007bq}
\bibinfo{author}{\bibfnamefont{A.~V.} \bibnamefont{Anisovich}}, \bibinfo{author}{\bibfnamefont{V.}~\bibnamefont{Kleber}}, \bibinfo{author}{\bibfnamefont{E.}~\bibnamefont{Klempt}}, \bibinfo{author}{\bibfnamefont{V.~A.} \bibnamefont{Nikonov}}, \bibinfo{author}{\bibfnamefont{A.~V.} \bibnamefont{Sarantsev}}, \bibnamefont{and} \bibinfo{author}{\bibfnamefont{U.}~\bibnamefont{Thoma}}, \bibinfo{journal}{Eur. Phys. J. A} \textbf{\bibinfo{volume}{34}}, \bibinfo{pages}{243} (\bibinfo{year}{2007}).

\bibitem[{\citenamefont{Wilkinson et~al.}(1981)}]{Wilkinson:1981jy}
\bibinfo{author}{\bibfnamefont{C.}~\bibnamefont{Wilkinson}} \bibnamefont{et~al.}, \bibinfo{journal}{Phys. Rev. Lett.} \textbf{\bibinfo{volume}{46}}, \bibinfo{pages}{803} (\bibinfo{year}{1981}).

\bibitem[{\citenamefont{Paterson et~al.}(2016)}]{CLAS:2016wrl}
\bibinfo{author}{\bibfnamefont{C.~A.} \bibnamefont{Paterson}} \bibnamefont{et~al.} (\bibinfo{collaboration}{CLAS}), \bibinfo{journal}{Phys. Rev. C} \textbf{\bibinfo{volume}{93}}, \bibinfo{pages}{065201} (\bibinfo{year}{2016}).

\bibitem[{\citenamefont{de~la Vaissiere et~al.}(1985)}]{delaVaissiere:1984xg}
\bibinfo{author}{\bibfnamefont{C.}~\bibnamefont{de~la Vaissiere}} \bibnamefont{et~al.}, \bibinfo{journal}{Phys. Rev. Lett.} \textbf{\bibinfo{volume}{54}}, \bibinfo{pages}{2071} (\bibinfo{year}{1985}), \bibinfo{note}{[Erratum: Phys.Rev.Lett. 55, 263 (1985)]}.

\bibitem[{\citenamefont{Goers et~al.}(1999)}]{SAPHIR:1999wfu}
\bibinfo{author}{\bibfnamefont{S.}~\bibnamefont{Goers}} \bibnamefont{et~al.} (\bibinfo{collaboration}{SAPHIR}), \bibinfo{journal}{Phys. Lett. B} \textbf{\bibinfo{volume}{464}}, \bibinfo{pages}{331} (\bibinfo{year}{1999}).

\bibitem[{\citenamefont{Ablikim et~al.}(2021)}]{BESIII:2020uqk}
\bibinfo{author}{\bibfnamefont{M.}~\bibnamefont{Ablikim}} \bibnamefont{et~al.} (\bibinfo{collaboration}{BESIII}), \bibinfo{journal}{Phys. Lett. B} \textbf{\bibinfo{volume}{814}}, \bibinfo{pages}{136110} (\bibinfo{year}{2021}).

\bibitem[{\citenamefont{Aduszkiewicz et~al.}(2016)}]{NA61SHINE:2015haq}
\bibinfo{author}{\bibfnamefont{A.}~\bibnamefont{Aduszkiewicz}} \bibnamefont{et~al.} (\bibinfo{collaboration}{NA61/SHINE}), \bibinfo{journal}{Eur. Phys. J. C} \textbf{\bibinfo{volume}{76}}, \bibinfo{pages}{198} (\bibinfo{year}{2016}).

\bibitem[{\citenamefont{Biagi et~al.}(1981)}]{Biagi:1981cu}
\bibinfo{author}{\bibfnamefont{S.~F.} \bibnamefont{Biagi}} \bibnamefont{et~al.}, \bibinfo{journal}{Z. Phys. C} \textbf{\bibinfo{volume}{9}}, \bibinfo{pages}{305} (\bibinfo{year}{1981}).

\bibitem[{\citenamefont{Dobbs et~al.}(2014)\citenamefont{Dobbs, Tomaradze, Xiao, Seth, and Bonvicini}}]{Dobbs:2014ifa}
\bibinfo{author}{\bibfnamefont{S.}~\bibnamefont{Dobbs}}, \bibinfo{author}{\bibfnamefont{A.}~\bibnamefont{Tomaradze}}, \bibinfo{author}{\bibfnamefont{T.}~\bibnamefont{Xiao}}, \bibinfo{author}{\bibfnamefont{K.~K.} \bibnamefont{Seth}}, \bibnamefont{and} \bibinfo{author}{\bibfnamefont{G.}~\bibnamefont{Bonvicini}}, \bibinfo{journal}{Phys. Lett. B} \textbf{\bibinfo{volume}{739}}, \bibinfo{pages}{90} (\bibinfo{year}{2014}).

\bibitem[{\citenamefont{Adamova et~al.}(2017)}]{ALICE:2017pgw}
\bibinfo{author}{\bibfnamefont{D.}~\bibnamefont{Adamova}} \bibnamefont{et~al.} (\bibinfo{collaboration}{ALICE}), \bibinfo{journal}{Eur. Phys. J. C} \textbf{\bibinfo{volume}{77}}, \bibinfo{pages}{389} (\bibinfo{year}{2017}).

\bibitem[{\citenamefont{Chien et~al.}(1966)\citenamefont{Chien, Lach, Sandweiss, Taft, Yeh, Oren, and Webster}}]{Chien:1966tr}
\bibinfo{author}{\bibfnamefont{C.~Y.} \bibnamefont{Chien}}, \bibinfo{author}{\bibfnamefont{J.}~\bibnamefont{Lach}}, \bibinfo{author}{\bibfnamefont{J.}~\bibnamefont{Sandweiss}}, \bibinfo{author}{\bibfnamefont{H.~D.} \bibnamefont{Taft}}, \bibinfo{author}{\bibfnamefont{N.}~\bibnamefont{Yeh}}, \bibinfo{author}{\bibfnamefont{Y.}~\bibnamefont{Oren}}, \bibnamefont{and} \bibinfo{author}{\bibfnamefont{M.}~\bibnamefont{Webster}}, \bibinfo{journal}{Phys. Rev.} \textbf{\bibinfo{volume}{152}}, \bibinfo{pages}{1171} (\bibinfo{year}{1966}).

\bibitem[{\citenamefont{Sarantsev et~al.}(2019)\citenamefont{Sarantsev, Matveev, Nikonov, Anisovich, Thoma, and Klempt}}]{Sarantsev:2019xxm}
\bibinfo{author}{\bibfnamefont{A.~V.} \bibnamefont{Sarantsev}}, \bibinfo{author}{\bibfnamefont{M.}~\bibnamefont{Matveev}}, \bibinfo{author}{\bibfnamefont{V.~A.} \bibnamefont{Nikonov}}, \bibinfo{author}{\bibfnamefont{A.~V.} \bibnamefont{Anisovich}}, \bibinfo{author}{\bibfnamefont{U.}~\bibnamefont{Thoma}}, \bibnamefont{and} \bibinfo{author}{\bibfnamefont{E.}~\bibnamefont{Klempt}}, \bibinfo{journal}{Eur. Phys. J. A} \textbf{\bibinfo{volume}{55}}, \bibinfo{pages}{180} (\bibinfo{year}{2019}).

\bibitem[{\citenamefont{Carmony et~al.}(1964)\citenamefont{Carmony, Pjerrou, Schlein, Slater, and Stork}}]{Carmony:1964zza}
\bibinfo{author}{\bibfnamefont{D.~D.} \bibnamefont{Carmony}}, \bibinfo{author}{\bibfnamefont{G.~M.} \bibnamefont{Pjerrou}}, \bibinfo{author}{\bibfnamefont{P.~E.} \bibnamefont{Schlein}}, \bibinfo{author}{\bibfnamefont{W.~E.} \bibnamefont{Slater}}, \bibnamefont{and} \bibinfo{author}{\bibfnamefont{D.~H.} \bibnamefont{Stork}}, \bibinfo{journal}{Phys. Rev. Lett.} \textbf{\bibinfo{volume}{12}}, \bibinfo{pages}{482} (\bibinfo{year}{1964}).

\bibitem[{\citenamefont{Beretvas et~al.}(1986)}]{Beretvas:1986km}
\bibinfo{author}{\bibfnamefont{A.}~\bibnamefont{Beretvas}} \bibnamefont{et~al.}, \bibinfo{journal}{Phys. Rev. D} \textbf{\bibinfo{volume}{34}}, \bibinfo{pages}{53} (\bibinfo{year}{1986}).

\bibitem[{\citenamefont{Bardadin-Otwinowska et~al.}(1975)}]{Bardadin-Otwinowska:1974lrc}
\bibinfo{author}{\bibfnamefont{M.}~\bibnamefont{Bardadin-Otwinowska}} \bibnamefont{et~al.}, \bibinfo{journal}{Nucl. Phys. B} \textbf{\bibinfo{volume}{90}}, \bibinfo{pages}{397} (\bibinfo{year}{1975}).

\bibitem[{\citenamefont{Ronniger and Metsch}(2011)}]{Ronniger:2011td}
\bibinfo{author}{\bibfnamefont{M.}~\bibnamefont{Ronniger}} \bibnamefont{and} \bibinfo{author}{\bibfnamefont{B.~C.} \bibnamefont{Metsch}}, \bibinfo{journal}{Eur. Phys. J. A} \textbf{\bibinfo{volume}{47}}, \bibinfo{pages}{162} (\bibinfo{year}{2011}).

\bibitem[{\citenamefont{Adamczewski-Musch et~al.}(2018)}]{HADES:2017njk}
\bibinfo{author}{\bibfnamefont{J.}~\bibnamefont{Adamczewski-Musch}} \bibnamefont{et~al.} (\bibinfo{collaboration}{HADES}), \bibinfo{journal}{Phys. Lett. B} \textbf{\bibinfo{volume}{781}}, \bibinfo{pages}{735} (\bibinfo{year}{2018}).

\bibitem[{\citenamefont{Crede}(2023)}]{Crede:2023ncq}
\bibinfo{author}{\bibfnamefont{V.}~\bibnamefont{Crede}} (\bibinfo{collaboration}{GlueX}), \bibinfo{journal}{Few Body Syst.} \textbf{\bibinfo{volume}{64}}, \bibinfo{pages}{32} (\bibinfo{year}{2023}).

\bibitem[{\citenamefont{Abazov et~al.}(2023)}]{PANDA:2023ljx}
\bibinfo{author}{\bibfnamefont{V.}~\bibnamefont{Abazov}} \bibnamefont{et~al.} (\bibinfo{collaboration}{PANDA}) (\bibinfo{year}{2023}).

\bibitem[{\citenamefont{Arellano and Adriazola}(2024)}]{Arellano:2024fym}
\bibinfo{author}{\bibfnamefont{H.~F.} \bibnamefont{Arellano}} \bibnamefont{and} \bibinfo{author}{\bibfnamefont{N.~A.} \bibnamefont{Adriazola}}, \bibinfo{journal}{Eur. Phys. J. A} \textbf{\bibinfo{volume}{60}}, \bibinfo{pages}{158} (\bibinfo{year}{2024}).

\bibitem[{\citenamefont{Nakamura and Jido}(2014)}]{Nakamura:2013boa}
\bibinfo{author}{\bibfnamefont{S.~X.} \bibnamefont{Nakamura}} \bibnamefont{and} \bibinfo{author}{\bibfnamefont{D.}~\bibnamefont{Jido}}, \bibinfo{journal}{PTEP} \textbf{\bibinfo{volume}{2014}}, \bibinfo{pages}{023D01} (\bibinfo{year}{2014}).

\bibitem[{\citenamefont{Van~Cauteren et~al.}(2004)\citenamefont{Van~Cauteren, Merten, Corthals, Janssen, Metsch, Petry, and Ryckebusch}}]{VanCauteren:2003hn}
\bibinfo{author}{\bibfnamefont{T.}~\bibnamefont{Van~Cauteren}}, \bibinfo{author}{\bibfnamefont{D.}~\bibnamefont{Merten}}, \bibinfo{author}{\bibfnamefont{T.}~\bibnamefont{Corthals}}, \bibinfo{author}{\bibfnamefont{S.}~\bibnamefont{Janssen}}, \bibinfo{author}{\bibfnamefont{B.}~\bibnamefont{Metsch}}, \bibinfo{author}{\bibfnamefont{H.~R.} \bibnamefont{Petry}}, \bibnamefont{and} \bibinfo{author}{\bibfnamefont{J.}~\bibnamefont{Ryckebusch}}, \bibinfo{journal}{Eur. Phys. J. A} \textbf{\bibinfo{volume}{20}}, \bibinfo{pages}{283} (\bibinfo{year}{2004}).

\bibitem[{\citenamefont{Jackson et~al.}(2015)\citenamefont{Jackson, Oh, Haberzettl, and Nakayama}}]{Jackson:2015dva}
\bibinfo{author}{\bibfnamefont{B.~C.} \bibnamefont{Jackson}}, \bibinfo{author}{\bibfnamefont{Y.}~\bibnamefont{Oh}}, \bibinfo{author}{\bibfnamefont{H.}~\bibnamefont{Haberzettl}}, \bibnamefont{and} \bibinfo{author}{\bibfnamefont{K.}~\bibnamefont{Nakayama}}, \bibinfo{journal}{Phys. Rev. C} \textbf{\bibinfo{volume}{91}}, \bibinfo{pages}{065208} (\bibinfo{year}{2015}).

\bibitem[{\citenamefont{Bernard et~al.}(1992)\citenamefont{Bernard, Kaiser, Kambor, and Meissner}}]{Bernard:1992xi}
\bibinfo{author}{\bibfnamefont{V.}~\bibnamefont{Bernard}}, \bibinfo{author}{\bibfnamefont{N.}~\bibnamefont{Kaiser}}, \bibinfo{author}{\bibfnamefont{J.}~\bibnamefont{Kambor}}, \bibnamefont{and} \bibinfo{author}{\bibfnamefont{U.~G.} \bibnamefont{Meissner}}, \bibinfo{journal}{Phys. Rev. D} \textbf{\bibinfo{volume}{46}}, \bibinfo{pages}{R2756} (\bibinfo{year}{1992}).

\bibitem[{\citenamefont{Lee}(1968)}]{Lee:1968ehl}
\bibinfo{author}{\bibfnamefont{B.~W.} \bibnamefont{Lee}}, \bibinfo{journal}{Phys. Rev.} \textbf{\bibinfo{volume}{170}}, \bibinfo{pages}{1359} (\bibinfo{year}{1968}).

\bibitem[{\citenamefont{Feldman et~al.}(1961)\citenamefont{Feldman, Matthews, and Salam}}]{Feldman:1961su}
\bibinfo{author}{\bibfnamefont{G.}~\bibnamefont{Feldman}}, \bibinfo{author}{\bibfnamefont{P.~T.} \bibnamefont{Matthews}}, \bibnamefont{and} \bibinfo{author}{\bibfnamefont{A.}~\bibnamefont{Salam}}, \bibinfo{journal}{Phys. Rev.} \textbf{\bibinfo{volume}{121}}, \bibinfo{pages}{302} (\bibinfo{year}{1961}).

\bibitem[{\citenamefont{Sibirtsev et~al.}(2006)\citenamefont{Sibirtsev, Haidenbauer, Hammer, and Meissner}}]{Sibirtsev:2006uy}
\bibinfo{author}{\bibfnamefont{A.}~\bibnamefont{Sibirtsev}}, \bibinfo{author}{\bibfnamefont{J.}~\bibnamefont{Haidenbauer}}, \bibinfo{author}{\bibfnamefont{H.~W.} \bibnamefont{Hammer}}, \bibnamefont{and} \bibinfo{author}{\bibfnamefont{U.~G.} \bibnamefont{Meissner}}, \bibinfo{journal}{Eur. Phys. J. A} \textbf{\bibinfo{volume}{29}}, \bibinfo{pages}{363} (\bibinfo{year}{2006}).

\bibitem[{\citenamefont{Ellis et~al.}(2007)\citenamefont{Ellis, Kotzinian, Naumov, and Sapozhnikov}}]{Ellis:2007ig}
\bibinfo{author}{\bibfnamefont{J.~R.} \bibnamefont{Ellis}}, \bibinfo{author}{\bibfnamefont{A.}~\bibnamefont{Kotzinian}}, \bibinfo{author}{\bibfnamefont{D.}~\bibnamefont{Naumov}}, \bibnamefont{and} \bibinfo{author}{\bibfnamefont{M.}~\bibnamefont{Sapozhnikov}}, \bibinfo{journal}{Eur. Phys. J. C} \textbf{\bibinfo{volume}{52}}, \bibinfo{pages}{283} (\bibinfo{year}{2007}).

\bibitem[{\citenamefont{Kim et~al.}(2013)\citenamefont{Kim, Nam, Hosaka, and Kim}}]{Kim:2012pz}
\bibinfo{author}{\bibfnamefont{S.-H.} \bibnamefont{Kim}}, \bibinfo{author}{\bibfnamefont{S.-i.} \bibnamefont{Nam}}, \bibinfo{author}{\bibfnamefont{A.}~\bibnamefont{Hosaka}}, \bibnamefont{and} \bibinfo{author}{\bibfnamefont{H.-C.} \bibnamefont{Kim}}, \bibinfo{journal}{Phys. Rev. D} \textbf{\bibinfo{volume}{88}}, \bibinfo{pages}{054012} (\bibinfo{year}{2013}).

\bibitem[{\citenamefont{Kaiser}(2005)}]{Kaiser:2005tu}
\bibinfo{author}{\bibfnamefont{N.}~\bibnamefont{Kaiser}}, \bibinfo{journal}{Phys. Rev. C} \textbf{\bibinfo{volume}{71}}, \bibinfo{pages}{068201} (\bibinfo{year}{2005}).

\bibitem[{\citenamefont{Ozpineci et~al.}(2005)\citenamefont{Ozpineci, Yakovlev, and Zamiralov}}]{Ozpineci:2003gm}
\bibinfo{author}{\bibfnamefont{A.}~\bibnamefont{Ozpineci}}, \bibinfo{author}{\bibfnamefont{S.~B.} \bibnamefont{Yakovlev}}, \bibnamefont{and} \bibinfo{author}{\bibfnamefont{V.~S.} \bibnamefont{Zamiralov}}, \bibinfo{journal}{Mod. Phys. Lett. A} \textbf{\bibinfo{volume}{20}}, \bibinfo{pages}{243} (\bibinfo{year}{2005}).

\bibitem[{\citenamefont{Zhong and Zhao}(2013)}]{Zhong:2013oqa}
\bibinfo{author}{\bibfnamefont{X.-H.} \bibnamefont{Zhong}} \bibnamefont{and} \bibinfo{author}{\bibfnamefont{Q.}~\bibnamefont{Zhao}}, \bibinfo{journal}{Phys. Rev. C} \textbf{\bibinfo{volume}{88}}, \bibinfo{pages}{015208} (\bibinfo{year}{2013}).

\bibitem[{\citenamefont{Quigg and Rosner}(1976)}]{Quigg:1976xs}
\bibinfo{author}{\bibfnamefont{C.}~\bibnamefont{Quigg}} \bibnamefont{and} \bibinfo{author}{\bibfnamefont{J.~L.} \bibnamefont{Rosner}}, \bibinfo{journal}{Phys. Rev. D} \textbf{\bibinfo{volume}{14}}, \bibinfo{pages}{160} (\bibinfo{year}{1976}).

\bibitem[{\citenamefont{Shi et~al.}(2023)\citenamefont{Shi, Gui, Liang, and Liu}}]{Shi:2023xfz}
\bibinfo{author}{\bibfnamefont{J.}~\bibnamefont{Shi}}, \bibinfo{author}{\bibfnamefont{L.-C.} \bibnamefont{Gui}}, \bibinfo{author}{\bibfnamefont{J.}~\bibnamefont{Liang}}, \bibnamefont{and} \bibinfo{author}{\bibfnamefont{G.}~\bibnamefont{Liu}} (\bibinfo{year}{2023}).

\bibitem[{\citenamefont{Yan et~al.}(2023)\citenamefont{Yan, Chen, and Xie}}]{Yan:2023yff}
\bibinfo{author}{\bibfnamefont{B.}~\bibnamefont{Yan}}, \bibinfo{author}{\bibfnamefont{C.}~\bibnamefont{Chen}}, \bibnamefont{and} \bibinfo{author}{\bibfnamefont{J.-J.} \bibnamefont{Xie}}, \bibinfo{journal}{Phys. Rev. D} \textbf{\bibinfo{volume}{107}}, \bibinfo{pages}{076008} (\bibinfo{year}{2023}).

\bibitem[{\citenamefont{Dalitz and Tuan}(1960)}]{Dalitz:1960du}
\bibinfo{author}{\bibfnamefont{R.~H.} \bibnamefont{Dalitz}} \bibnamefont{and} \bibinfo{author}{\bibfnamefont{S.~F.} \bibnamefont{Tuan}}, \bibinfo{journal}{Annals Phys.} \textbf{\bibinfo{volume}{10}}, \bibinfo{pages}{307} (\bibinfo{year}{1960}).

\bibitem[{\citenamefont{Dalitz and Tuan}(1959)}]{Dalitz:1959dn}
\bibinfo{author}{\bibfnamefont{R.~H.} \bibnamefont{Dalitz}} \bibnamefont{and} \bibinfo{author}{\bibfnamefont{S.~F.} \bibnamefont{Tuan}}, \bibinfo{journal}{Phys. Rev. Lett.} \textbf{\bibinfo{volume}{2}}, \bibinfo{pages}{425} (\bibinfo{year}{1959}).

\bibitem[{\citenamefont{Kaiser et~al.}(1995{\natexlab{a}})\citenamefont{Kaiser, Siegel, and Weise}}]{Kaiser:1995cy}
\bibinfo{author}{\bibfnamefont{N.}~\bibnamefont{Kaiser}}, \bibinfo{author}{\bibfnamefont{P.~B.} \bibnamefont{Siegel}}, \bibnamefont{and} \bibinfo{author}{\bibfnamefont{W.}~\bibnamefont{Weise}}, \bibinfo{journal}{Phys. Lett. B} \textbf{\bibinfo{volume}{362}}, \bibinfo{pages}{23} (\bibinfo{year}{1995}{\natexlab{a}}).

\bibitem[{\citenamefont{Kaiser et~al.}(1995{\natexlab{b}})\citenamefont{Kaiser, Siegel, and Weise}}]{Kaiser:1995eg}
\bibinfo{author}{\bibfnamefont{N.}~\bibnamefont{Kaiser}}, \bibinfo{author}{\bibfnamefont{P.~B.} \bibnamefont{Siegel}}, \bibnamefont{and} \bibinfo{author}{\bibfnamefont{W.}~\bibnamefont{Weise}}, \bibinfo{journal}{Nucl. Phys. A} \textbf{\bibinfo{volume}{594}}, \bibinfo{pages}{325} (\bibinfo{year}{1995}{\natexlab{b}}).

\bibitem[{\citenamefont{Oset and Ramos}(1998)}]{Oset:1997it}
\bibinfo{author}{\bibfnamefont{E.}~\bibnamefont{Oset}} \bibnamefont{and} \bibinfo{author}{\bibfnamefont{A.}~\bibnamefont{Ramos}}, \bibinfo{journal}{Nucl. Phys. A} \textbf{\bibinfo{volume}{635}}, \bibinfo{pages}{99} (\bibinfo{year}{1998}).

\bibitem[{\citenamefont{Oller et~al.}(2000)\citenamefont{Oller, Oset, and Ramos}}]{Oller:2000ma}
\bibinfo{author}{\bibfnamefont{J.~A.} \bibnamefont{Oller}}, \bibinfo{author}{\bibfnamefont{E.}~\bibnamefont{Oset}}, \bibnamefont{and} \bibinfo{author}{\bibfnamefont{A.}~\bibnamefont{Ramos}}, \bibinfo{journal}{Prog. Part. Nucl. Phys.} \textbf{\bibinfo{volume}{45}}, \bibinfo{pages}{157} (\bibinfo{year}{2000}).

\bibitem[{\citenamefont{Ecker}(1995)}]{Ecker:1994gg}
\bibinfo{author}{\bibfnamefont{G.}~\bibnamefont{Ecker}}, \bibinfo{journal}{Prog. Part. Nucl. Phys.} \textbf{\bibinfo{volume}{35}}, \bibinfo{pages}{1} (\bibinfo{year}{1995}).

\bibitem[{\citenamefont{Bernard et~al.}(1995)\citenamefont{Bernard, Kaiser, and Meissner}}]{Bernard:1995dp}
\bibinfo{author}{\bibfnamefont{V.}~\bibnamefont{Bernard}}, \bibinfo{author}{\bibfnamefont{N.}~\bibnamefont{Kaiser}}, \bibnamefont{and} \bibinfo{author}{\bibfnamefont{U.-G.} \bibnamefont{Meissner}}, \bibinfo{journal}{Int. J. Mod. Phys. E} \textbf{\bibinfo{volume}{4}}, \bibinfo{pages}{193} (\bibinfo{year}{1995}).

\bibitem[{\citenamefont{Oller and Meissner}(2001)}]{Oller:2000fj}
\bibinfo{author}{\bibfnamefont{J.~A.} \bibnamefont{Oller}} \bibnamefont{and} \bibinfo{author}{\bibfnamefont{U.~G.} \bibnamefont{Meissner}}, \bibinfo{journal}{Phys. Lett. B} \textbf{\bibinfo{volume}{500}}, \bibinfo{pages}{263} (\bibinfo{year}{2001}).

\bibitem[{\citenamefont{Jido et~al.}(2003)\citenamefont{Jido, Oller, Oset, Ramos, and Meissner}}]{Jido:2003cb}
\bibinfo{author}{\bibfnamefont{D.}~\bibnamefont{Jido}}, \bibinfo{author}{\bibfnamefont{J.~A.} \bibnamefont{Oller}}, \bibinfo{author}{\bibfnamefont{E.}~\bibnamefont{Oset}}, \bibinfo{author}{\bibfnamefont{A.}~\bibnamefont{Ramos}}, \bibnamefont{and} \bibinfo{author}{\bibfnamefont{U.~G.} \bibnamefont{Meissner}}, \bibinfo{journal}{Nucl. Phys. A} \textbf{\bibinfo{volume}{725}}, \bibinfo{pages}{181} (\bibinfo{year}{2003}).

\bibitem[{\citenamefont{Navas et~al.}(2024)}]{ParticleDataGroup:2024cfk}
\bibinfo{author}{\bibfnamefont{S.}~\bibnamefont{Navas}} \bibnamefont{et~al.} (\bibinfo{collaboration}{Particle Data Group}), \bibinfo{journal}{Phys. Rev. D} \textbf{\bibinfo{volume}{110}}, \bibinfo{pages}{030001} (\bibinfo{year}{2024}).

\bibitem[{\citenamefont{Oset and Ramos}(2010{\natexlab{a}})}]{Oset:2010tof}
\bibinfo{author}{\bibfnamefont{E.}~\bibnamefont{Oset}} \bibnamefont{and} \bibinfo{author}{\bibfnamefont{A.}~\bibnamefont{Ramos}}, \bibinfo{journal}{Eur. Phys. J. A} \textbf{\bibinfo{volume}{44}}, \bibinfo{pages}{445} (\bibinfo{year}{2010}{\natexlab{a}}).

\bibitem[{\citenamefont{Lutz and Kolomeitsev}(2002)}]{Lutz:2001yb}
\bibinfo{author}{\bibfnamefont{M.~F.~M.} \bibnamefont{Lutz}} \bibnamefont{and} \bibinfo{author}{\bibfnamefont{E.~E.} \bibnamefont{Kolomeitsev}}, \bibinfo{journal}{Nucl. Phys. A} \textbf{\bibinfo{volume}{700}}, \bibinfo{pages}{193} (\bibinfo{year}{2002}).

\bibitem[{\citenamefont{Garcia-Recio et~al.}(2003)\citenamefont{Garcia-Recio, Nieves, Ruiz~Arriola, and Vicente~Vacas}}]{Garcia-Recio:2002yxy}
\bibinfo{author}{\bibfnamefont{C.}~\bibnamefont{Garcia-Recio}}, \bibinfo{author}{\bibfnamefont{J.}~\bibnamefont{Nieves}}, \bibinfo{author}{\bibfnamefont{E.}~\bibnamefont{Ruiz~Arriola}}, \bibnamefont{and} \bibinfo{author}{\bibfnamefont{M.~J.} \bibnamefont{Vicente~Vacas}}, \bibinfo{journal}{Phys. Rev. D} \textbf{\bibinfo{volume}{67}}, \bibinfo{pages}{076009} (\bibinfo{year}{2003}).

\bibitem[{\citenamefont{Magas et~al.}(2005)\citenamefont{Magas, Oset, and Ramos}}]{Magas:2005vu}
\bibinfo{author}{\bibfnamefont{V.~K.} \bibnamefont{Magas}}, \bibinfo{author}{\bibfnamefont{E.}~\bibnamefont{Oset}}, \bibnamefont{and} \bibinfo{author}{\bibfnamefont{A.}~\bibnamefont{Ramos}}, \bibinfo{journal}{Phys. Rev. Lett.} \textbf{\bibinfo{volume}{95}}, \bibinfo{pages}{052301} (\bibinfo{year}{2005}).

\bibitem[{\citenamefont{Ikeda et~al.}(2012)\citenamefont{Ikeda, Hyodo, and Weise}}]{Ikeda:2012au}
\bibinfo{author}{\bibfnamefont{Y.}~\bibnamefont{Ikeda}}, \bibinfo{author}{\bibfnamefont{T.}~\bibnamefont{Hyodo}}, \bibnamefont{and} \bibinfo{author}{\bibfnamefont{W.}~\bibnamefont{Weise}}, \bibinfo{journal}{Nucl. Phys. A} \textbf{\bibinfo{volume}{881}}, \bibinfo{pages}{98} (\bibinfo{year}{2012}).

\bibitem[{\citenamefont{Guo and Oller}(2013)}]{Guo:2012vv}
\bibinfo{author}{\bibfnamefont{Z.-H.} \bibnamefont{Guo}} \bibnamefont{and} \bibinfo{author}{\bibfnamefont{J.~A.} \bibnamefont{Oller}}, \bibinfo{journal}{Phys. Rev. C} \textbf{\bibinfo{volume}{87}}, \bibinfo{pages}{035202} (\bibinfo{year}{2013}).

\bibitem[{\citenamefont{Mai and Mei\ss{}ner}(2015)}]{Mai:2014xna}
\bibinfo{author}{\bibfnamefont{M.}~\bibnamefont{Mai}} \bibnamefont{and} \bibinfo{author}{\bibfnamefont{U.-G.} \bibnamefont{Mei\ss{}ner}}, \bibinfo{journal}{Eur. Phys. J. A} \textbf{\bibinfo{volume}{51}}, \bibinfo{pages}{30} (\bibinfo{year}{2015}).

\bibitem[{\citenamefont{Roca and Oset}(2013{\natexlab{a}})}]{Roca:2013av}
\bibinfo{author}{\bibfnamefont{L.}~\bibnamefont{Roca}} \bibnamefont{and} \bibinfo{author}{\bibfnamefont{E.}~\bibnamefont{Oset}}, \bibinfo{journal}{Phys. Rev. C} \textbf{\bibinfo{volume}{87}}, \bibinfo{pages}{055201} (\bibinfo{year}{2013}{\natexlab{a}}).

\bibitem[{\citenamefont{Roca and Oset}(2013{\natexlab{b}})}]{Roca:2013cca}
\bibinfo{author}{\bibfnamefont{L.}~\bibnamefont{Roca}} \bibnamefont{and} \bibinfo{author}{\bibfnamefont{E.}~\bibnamefont{Oset}}, \bibinfo{journal}{Phys. Rev. C} \textbf{\bibinfo{volume}{88}}, \bibinfo{pages}{055206} (\bibinfo{year}{2013}{\natexlab{b}}).

\bibitem[{\citenamefont{Ciepl\'y et~al.}(2016)\citenamefont{Ciepl\'y, Mai, Mei\ss{}ner, and Smejkal}}]{Cieply:2016jby}
\bibinfo{author}{\bibfnamefont{A.}~\bibnamefont{Ciepl\'y}}, \bibinfo{author}{\bibfnamefont{M.}~\bibnamefont{Mai}}, \bibinfo{author}{\bibfnamefont{U.-G.} \bibnamefont{Mei\ss{}ner}}, \bibnamefont{and} \bibinfo{author}{\bibfnamefont{J.}~\bibnamefont{Smejkal}}, \bibinfo{journal}{Nucl. Phys. A} \textbf{\bibinfo{volume}{954}}, \bibinfo{pages}{17} (\bibinfo{year}{2016}).

\bibitem[{\citenamefont{Cieply and Smejkal}(2012)}]{Cieply:2011nq}
\bibinfo{author}{\bibfnamefont{A.}~\bibnamefont{Cieply}} \bibnamefont{and} \bibinfo{author}{\bibfnamefont{J.}~\bibnamefont{Smejkal}}, \bibinfo{journal}{Nucl. Phys. A} \textbf{\bibinfo{volume}{881}}, \bibinfo{pages}{115} (\bibinfo{year}{2012}).

\bibitem[{\citenamefont{Kamiya et~al.}(2016)\citenamefont{Kamiya, Miyahara, Ohnishi, Ikeda, Hyodo, Oset, and Weise}}]{Kamiya:2016jqc}
\bibinfo{author}{\bibfnamefont{Y.}~\bibnamefont{Kamiya}}, \bibinfo{author}{\bibfnamefont{K.}~\bibnamefont{Miyahara}}, \bibinfo{author}{\bibfnamefont{S.}~\bibnamefont{Ohnishi}}, \bibinfo{author}{\bibfnamefont{Y.}~\bibnamefont{Ikeda}}, \bibinfo{author}{\bibfnamefont{T.}~\bibnamefont{Hyodo}}, \bibinfo{author}{\bibfnamefont{E.}~\bibnamefont{Oset}}, \bibnamefont{and} \bibinfo{author}{\bibfnamefont{W.}~\bibnamefont{Weise}}, \bibinfo{journal}{Nucl. Phys. A} \textbf{\bibinfo{volume}{954}}, \bibinfo{pages}{41} (\bibinfo{year}{2016}).

\bibitem[{\citenamefont{Hyodo and Weise}(2008)}]{Hyodo:2007jq}
\bibinfo{author}{\bibfnamefont{T.}~\bibnamefont{Hyodo}} \bibnamefont{and} \bibinfo{author}{\bibfnamefont{W.}~\bibnamefont{Weise}}, \bibinfo{journal}{Phys. Rev. C} \textbf{\bibinfo{volume}{77}}, \bibinfo{pages}{035204} (\bibinfo{year}{2008}).

\bibitem[{\citenamefont{R\'evai}(2018)}]{Revai:2017isg}
\bibinfo{author}{\bibfnamefont{J.}~\bibnamefont{R\'evai}}, \bibinfo{journal}{Few Body Syst.} \textbf{\bibinfo{volume}{59}}, \bibinfo{pages}{49} (\bibinfo{year}{2018}).

\bibitem[{\citenamefont{Bruns and Ciepl\'y}(2020)}]{Bruns:2019bwg}
\bibinfo{author}{\bibfnamefont{P.~C.} \bibnamefont{Bruns}} \bibnamefont{and} \bibinfo{author}{\bibfnamefont{A.}~\bibnamefont{Ciepl\'y}}, \bibinfo{journal}{Nucl. Phys. A} \textbf{\bibinfo{volume}{996}}, \bibinfo{pages}{121702} (\bibinfo{year}{2020}).

\bibitem[{\citenamefont{Miyahara and Hyodo}(2018)}]{Miyahara:2018lud}
\bibinfo{author}{\bibfnamefont{K.}~\bibnamefont{Miyahara}} \bibnamefont{and} \bibinfo{author}{\bibfnamefont{T.}~\bibnamefont{Hyodo}}, \bibinfo{journal}{Phys. Rev. C} \textbf{\bibinfo{volume}{98}}, \bibinfo{pages}{025202} (\bibinfo{year}{2018}).

\bibitem[{\citenamefont{Hyodo and Jido}(2012)}]{Hyodo:2011ur}
\bibinfo{author}{\bibfnamefont{T.}~\bibnamefont{Hyodo}} \bibnamefont{and} \bibinfo{author}{\bibfnamefont{D.}~\bibnamefont{Jido}}, \bibinfo{journal}{Prog. Part. Nucl. Phys.} \textbf{\bibinfo{volume}{67}}, \bibinfo{pages}{55} (\bibinfo{year}{2012}).

\bibitem[{\citenamefont{Mei\ss{}ner}(2020)}]{Meissner:2020khl}
\bibinfo{author}{\bibfnamefont{U.-G.} \bibnamefont{Mei\ss{}ner}}, \bibinfo{journal}{Symmetry} \textbf{\bibinfo{volume}{12}}, \bibinfo{pages}{981} (\bibinfo{year}{2020}).

\bibitem[{\citenamefont{Moriya et~al.}(2013{\natexlab{a}})}]{CLAS:2013rjt}
\bibinfo{author}{\bibfnamefont{K.}~\bibnamefont{Moriya}} \bibnamefont{et~al.} (\bibinfo{collaboration}{CLAS}), \bibinfo{journal}{Phys. Rev. C} \textbf{\bibinfo{volume}{87}}, \bibinfo{pages}{035206} (\bibinfo{year}{2013}{\natexlab{a}}).

\bibitem[{\citenamefont{Moriya et~al.}(2013{\natexlab{b}})}]{CLAS:2013rxx}
\bibinfo{author}{\bibfnamefont{K.}~\bibnamefont{Moriya}} \bibnamefont{et~al.} (\bibinfo{collaboration}{CLAS}), \bibinfo{journal}{Phys. Rev. C} \textbf{\bibinfo{volume}{88}}, \bibinfo{pages}{045201} (\bibinfo{year}{2013}{\natexlab{b}}), \bibinfo{note}{[Addendum: Phys.Rev.C 88, 049902 (2013)]}.

\bibitem[{\citenamefont{Schumacher and Moriya}(2013)}]{Schumacher:2013vma}
\bibinfo{author}{\bibfnamefont{R.~A.} \bibnamefont{Schumacher}} \bibnamefont{and} \bibinfo{author}{\bibfnamefont{K.}~\bibnamefont{Moriya}}, \bibinfo{journal}{Nucl. Phys. A} \textbf{\bibinfo{volume}{914}}, \bibinfo{pages}{51} (\bibinfo{year}{2013}).

\bibitem[{\citenamefont{Bando et~al.}(1985)\citenamefont{Bando, Kugo, Uehara, Yamawaki, and Yanagida}}]{BandoLHG1}
\bibinfo{author}{\bibfnamefont{M.}~\bibnamefont{Bando}}, \bibinfo{author}{\bibfnamefont{T.}~\bibnamefont{Kugo}}, \bibinfo{author}{\bibfnamefont{S.}~\bibnamefont{Uehara}}, \bibinfo{author}{\bibfnamefont{K.}~\bibnamefont{Yamawaki}}, \bibnamefont{and} \bibinfo{author}{\bibfnamefont{T.}~\bibnamefont{Yanagida}}, \bibinfo{journal}{Phys. Rev. Lett.} \textbf{\bibinfo{volume}{54}}, \bibinfo{pages}{1215} (\bibinfo{year}{1985}).

\bibitem[{\citenamefont{Bando et~al.}(1988)\citenamefont{Bando, Kugo, and Yamawaki}}]{BandoLHG2}
\bibinfo{author}{\bibfnamefont{M.}~\bibnamefont{Bando}}, \bibinfo{author}{\bibfnamefont{T.}~\bibnamefont{Kugo}}, \bibnamefont{and} \bibinfo{author}{\bibfnamefont{K.}~\bibnamefont{Yamawaki}}, \bibinfo{journal}{Phys. Rept.} \textbf{\bibinfo{volume}{164}}, \bibinfo{pages}{217} (\bibinfo{year}{1988}).

\bibitem[{\citenamefont{Meissner}(1988)}]{UlfLHG}
\bibinfo{author}{\bibfnamefont{U.~G.} \bibnamefont{Meissner}}, \bibinfo{journal}{Phys. Rept.} \textbf{\bibinfo{volume}{161}}, \bibinfo{pages}{213} (\bibinfo{year}{1988}).

\bibitem[{\citenamefont{Nagahiro et~al.}(2009)\citenamefont{Nagahiro, Roca, Hosaka, and Oset}}]{NagahiroVDM}
\bibinfo{author}{\bibfnamefont{H.}~\bibnamefont{Nagahiro}}, \bibinfo{author}{\bibfnamefont{L.}~\bibnamefont{Roca}}, \bibinfo{author}{\bibfnamefont{A.}~\bibnamefont{Hosaka}}, \bibnamefont{and} \bibinfo{author}{\bibfnamefont{E.}~\bibnamefont{Oset}}, \bibinfo{journal}{Phys. Rev. D} \textbf{\bibinfo{volume}{79}}, \bibinfo{pages}{014015} (\bibinfo{year}{2009}), \eprint{0809.0943}.

\bibitem[{\citenamefont{Bramon et~al.}(1992)\citenamefont{Bramon, Grau, and Pancheri}}]{Bramon92}
\bibinfo{author}{\bibfnamefont{A.}~\bibnamefont{Bramon}}, \bibinfo{author}{\bibfnamefont{A.}~\bibnamefont{Grau}}, \bibnamefont{and} \bibinfo{author}{\bibfnamefont{G.}~\bibnamefont{Pancheri}}, \bibinfo{journal}{Phys. Lett. B} \textbf{\bibinfo{volume}{283}}, \bibinfo{pages}{416} (\bibinfo{year}{1992}).

\bibitem[{\citenamefont{Oset et~al.}(2003)\citenamefont{Oset, Pelaez, and Roca}}]{Oset:2002sh}
\bibinfo{author}{\bibfnamefont{E.}~\bibnamefont{Oset}}, \bibinfo{author}{\bibfnamefont{J.~R.} \bibnamefont{Pelaez}}, \bibnamefont{and} \bibinfo{author}{\bibfnamefont{L.}~\bibnamefont{Roca}}, \bibinfo{journal}{Phys. Rev. D} \textbf{\bibinfo{volume}{67}}, \bibinfo{pages}{073013} (\bibinfo{year}{2003}).

\bibitem[{\citenamefont{Oset et~al.}(2008)\citenamefont{Oset, Pelaez, and Roca}}]{Oset:2008hp}
\bibinfo{author}{\bibfnamefont{E.}~\bibnamefont{Oset}}, \bibinfo{author}{\bibfnamefont{J.~R.} \bibnamefont{Pelaez}}, \bibnamefont{and} \bibinfo{author}{\bibfnamefont{L.}~\bibnamefont{Roca}}, \bibinfo{journal}{Phys. Rev. D} \textbf{\bibinfo{volume}{77}}, \bibinfo{pages}{073001} (\bibinfo{year}{2008}).

\bibitem[{\citenamefont{Oset and Ramos}(2010{\natexlab{b}})}]{RamosKN}
\bibinfo{author}{\bibfnamefont{E.}~\bibnamefont{Oset}} \bibnamefont{and} \bibinfo{author}{\bibfnamefont{A.}~\bibnamefont{Ramos}}, \bibinfo{journal}{Eur. Phys. J. A} \textbf{\bibinfo{volume}{44}}, \bibinfo{pages}{445} (\bibinfo{year}{2010}{\natexlab{b}}).

\bibitem[{\citenamefont{Garzon and Oset}(2012)}]{GarzonKN}
\bibinfo{author}{\bibfnamefont{E.~J.} \bibnamefont{Garzon}} \bibnamefont{and} \bibinfo{author}{\bibfnamefont{E.}~\bibnamefont{Oset}}, \bibinfo{journal}{Eur. Phys. J. A} \textbf{\bibinfo{volume}{48}}, \bibinfo{pages}{5} (\bibinfo{year}{2012}).

\bibitem[{\citenamefont{Mandl and Shaw}(1985)}]{Mandl}
\bibinfo{author}{\bibfnamefont{F.}~\bibnamefont{Mandl}} \bibnamefont{and} \bibinfo{author}{\bibfnamefont{G.}~\bibnamefont{Shaw}}, \emph{\bibinfo{title}{{QUANTUM FIELD THEORY}}} (\bibinfo{year}{1985}).

\bibitem[{\citenamefont{Gamermann et~al.}(2010)\citenamefont{Gamermann, Nieves, Oset, and Ruiz~Arriola}}]{danijuan}
\bibinfo{author}{\bibfnamefont{D.}~\bibnamefont{Gamermann}}, \bibinfo{author}{\bibfnamefont{J.}~\bibnamefont{Nieves}}, \bibinfo{author}{\bibfnamefont{E.}~\bibnamefont{Oset}}, \bibnamefont{and} \bibinfo{author}{\bibfnamefont{E.}~\bibnamefont{Ruiz~Arriola}}, \bibinfo{journal}{Phys. Rev. D} \textbf{\bibinfo{volume}{81}}, \bibinfo{pages}{014029} (\bibinfo{year}{2010}).

\end{thebibliography}
\end{document}